\documentclass[PAPER, cernpreprint, USenglish, texmf, orcidlogo]{atlasdoc}
\usepackage{atlaspackage}
\usepackage{atlasbiblatex}

\usepackage{makecell}
\usepackage{atlasphysics}
\graphicspath{{logos/}{figures/}}

\usepackage{ANA-SOFT-2023-01-PAPER-defs}

\AtlasTitle{An implementation of neural simulation-based inference for parameter estimation in ATLAS}

\AtlasAbstract{
Neural simulation-based inference is a powerful class of machine-learning-based methods for statistical inference that naturally handles high-dimensional parameter estimation without the need to bin data into low-dimensional summary histograms. Such methods are promising for a range of measurements, including at the Large Hadron Collider, where no single observable may be optimal to scan over the entire theoretical phase space under consideration, or where binning data into histograms could result in a loss of sensitivity. This work develops a neural simulation-based inference framework for statistical inference, using neural networks to estimate probability density ratios, which enables the application to a full-scale analysis. It incorporates a large number of systematic uncertainties, quantifies the uncertainty due to the finite number of events in training samples, develops a method to construct confidence intervals, and demonstrates a series of intermediate diagnostic checks that can be performed to validate the robustness of the method. As an example, the power and feasibility of the method are assessed on simulated data for a simplified version of an off-shell Higgs boson couplings measurement in the four-lepton final states. This approach represents an extension to the standard statistical methodology used by the experiments at the Large Hadron Collider, and can benefit many physics analyses.
}

\AtlasRefCode{SOFT-2023-01}

\PreprintIdNumber{CERN-EP-2024-305}

\arXivId{2412.01600}

\AtlasJournal{Rep. Prog. Phys.}
\AtlasJournalRef{Rep. Prog. Phys. 88 (2025) 067801}
\AtlasDOI{10.1088/1361-6633/add370}

\hypersetup{pdftitle={ATLAS document},pdfauthor={The ATLAS Collaboration}}

\begin{document}

\maketitle

\tableofcontents

\section{Introduction}

% The next lines are included from the .//./sections/Intro.tex input file
Precision measurements of theoretical parameters are a core element of the scientific program of experiments at the Large Hadron Collider (LHC). Such measurements typically rely on the method of maximum likelihood, which assesses the likelihood of the observed data for a range of parameter values~\cite{cowan1998statistical,Cranmer:2014lly}. While the likelihood cannot be calculated analytically, it can be estimated from high-fidelity simulations of data under varying parameter values~\cite{doi:10.1073/pnas.1912789117}. As the space of observational measurements grows to higher dimensionality, density estimation becomes challenging and is often preceded by data reduction, which compresses the relevant information into a low-dimensional summary statistic, often a single observable, allowing for simple probability density estimation methods, such as histogramming, to be used~\cite{cowan1998statistical}. While significant effort goes into both the design of this summary observable and the choice of histogram binning, these simplifications may nonetheless result in a loss of sensitivity. This is of particular concern for problems where the differential cross-sections and the observed kinematic distributions of different physics processes have a non-linear dependence on the parameter of interest (POI). In non-linear cases, no single observable may contain all the information required to maximize the sensitivity of the analysis over the full range of the theory parameter under consideration~\cite{Brehmer:2018eca,Ghosh_jrjc_proc,Butter:2021rvz}. Examples of analyses for which this can have significant consequences for the sensitivity include the off-shell Higgs boson production measurements and effective field theory (EFT) measurements, where quantum interference introduces a non-linear dependence on the parameter of interest.

While histograms cannot effectively scale to high dimensions, neural networks (NN) were shown to perform high-dimensional, unbinned density estimation in the context of parameter estimation at the LHC~\cite{Cranmer:2015bka,Brehmer:2018eca,Brehmer:2019xox,Heinrich:2022qlq,Heinrich:2023bmt} without the need to collapse information to a single observable. Referred to as \textit{neural simulation-based inference (NSBI)}, these methods can dramatically enhance sensitivity in analyses where the simplifications coming from using histograms of a single summary statistics are unwarranted.

NSBI techniques are of interest to a wide range of scientific fields for parameter estimation in cases where likelihoods are either intractable or computationally expensive to evaluate. When high-fidelity simulators can provide samples drawn from these likelihoods, NNs are capable of learning the underlying density of these simulated samples and can be used to approximate a likelihood ratio~\cite{Cranmer:2015bka}, the likelihood itself~\cite{papamakarios2021normalizingflowsprobabilisticmodeling}, or, in the context of Bayesian inference, the posterior~\cite{papamakarios2018fastepsilonfreeinferencesimulation}. These techniques have several potential applications in the physical sciences~\cite{doi:10.1073/pnas.1912789117,Brehmer:2018eca} and can be used, for instance, to study galaxy clustering~\cite{Modi:2023llw}, probe the interior of neutron stars from telescope data~\cite{Brandes:2024vhw}, probe the equation of state of neutral stars~\cite{Soma:2022vbb}, explore the QCD phase structure~\cite{Zhou:2023pti}, or analyze data from gravitational wave detectors~\cite{Dax:2021tsq}. A comprehensive summary of the use of NSBI methods in sciences can be found in \Refn{\cite{doi:10.1073/pnas.1912789117}}.

For the application of NSBI to analysis of experimental particle physics data, crucial questions remain unanswered. How can a large number of nuisance parameters (NP) be incorporated? How can the uncertainty from limited Monte-Carlo (MC) simulated samples be quantified? Can NN produce robust likelihood ratios and confidence intervals when applied in a realistic experimental context, and how can their reliability be effectively tested? This paper answers these questions and, thereby, enables the construction of a complete NSBI framework, together with diagnostic tools to address these questions.

The developed framework is an extension of an established statistical method at the LHC~\cite{cowan1998statistical,Cranmer:2014lly}, to be usable with unbinned, multidimensional data, where NNs are used to estimate likelihood ratios between hypotheses. It accounts for both linear and non-linear dependence of physics observables as a function of theory parameters, which is crucial to building optimal test statistic functions. The challenge of \textit{model misspecification} in NSBI, where the simulations have systematic differences from real data, is addressed by the introduction of NPs representing systematic uncertainties, and by testing the modeling of these systematic uncertainties themselves, in a similar way to how it is done for traditional analysis techniques in particle physics~\cite{Cranmer:1456844}. This method can leverage an analytical factorization of contributions from different physics processes to the full likelihood, which is possible in many analyses at the LHC\@. To show the power and applicability of the method, an example use case is described using samples describing the gluon fusion processes simulated for the ATLAS off-shell Higgs boson production measurement in the four-lepton final states~\cite{HIGG-2018-32}. The example will be compared to an equivalent \textit{histogram-based analysis}, defined as one where the likelihood ratio is given by a multinomial distribution based on the histogram of a single observable. The improvement in sensitivity compared to a histogram-based analysis, while accounting for systematic uncertainties, comes from the optimization of the analysis over the entire range of the theory parameter, which cannot be achieved with the use of only a single observable for all regions of the theory space, and from the unbinned nature of the method. The robustness tests that are needed to build a reliable likelihood ratio model using NNs are also shown.

Recently, the ATLAS experiment has also demonstrated the ability to unfold differential cross-sections in high-dimensional observable space and without binning~\cite{STDM-2020-17} using a technique that also relies on the ability of classifiers to estimate probability density ratios. The goals of unfolding are different from parameter estimation from experimental data, and the two approaches are complementary.

Since this paper builds upon established statistical methods at the LHC, it focuses on the tools and concepts necessary to extend these to a high-dimensional and unbinned NSBI analysis. The paper is organized as follows. \Sect{\ref{sec:NSBI}} reviews the concepts of NSBI and modifications that are developed for a practical application at the LHC\@. \Sect{\ref{sec:ExampleUseCase}} introduces the context of the off-shell Higgs boson production measurement, which is the example analysis used to demonstrate the method developed. \Sect{\ref{sec:diagnostics}} then describes the diagnostic tools used to validate the trained models, \Sect{\ref{sec:systematics}} extends the method to incorporate systematic uncertainties, \Sect{\ref{sec:NeymanConstruction}} describes how to build confidence intervals for NSBI, and \Sect{\ref{sec:resultComparison}} assesses the gain in sensitivity. The conclusion is presented in \Sect{\ref{sec:conclusion}} with a discussion of opportunities and challenges.

% End of text imported from the .//./sections/Intro.tex input file

\section{Neural simulation-based inference}
\label{sec:NSBI}

% The next lines are included from the .//./sections/NSBI.tex input file
This section reviews the core principles of classifier-based NSBI and discusses a framework in which the method can be made robust and numerically stable. NPs are introduced into this framework in \Sect{\ref{sec:systematics}}.

\subsection{Classifiers as probability density ratio estimators}

NN classifiers can be used to discriminate between two hypotheses $\mu_0$ and $\mu_1$ by minimizing the binary cross-entropy loss function,

\begin{equation}
\mathcal{L}[s] = -\frac{1}{(\sum_{i=1}^N w_i)}\sum_{i=1}^N w_i \cdot[y_i\log s(x_i) + (1-y_i)\log (1-s(x_i))],
\end{equation}
where the sum is over $N$ events $i$ sampled from probability density functions $p(x_i|\mu_0)$ or $p(x_i|\mu_1)$ with weights $w_i$ and assigned labels $y_i=0$ or $y_i=1$, respectively, and $s(x_i)$ is the classifier decision function. Each event $i$ is described by a vector of observables $x_i$.

The optimal decision function (in the infinite sample limit, i.e., as $N\to \infty$), which minimizes the binary cross entropy function, is given by~\cite{Cranmer:2015bka,hastie01statisticallearning}

\begin{equation}
s(x_i) = \frac{p(x_i|\mu_1)\cdot \nu(\mu_1)}{p(x_i|\mu_0)\cdot \nu(\mu_0) +  p(x_i|\mu_1)\cdot \nu(\mu_1)},
\end{equation}
where $\nu(\mu_0)$ and $\nu(\mu_1)$ are the expected number of events for each hypothesis.

In high-energy physics, training datasets are usually taken from MC simulated samples generated according to the two hypotheses. These simulated events are often weighted, and the weights may take negative values. Typically, the weights are scaled to perform the training with \textit{balanced samples}, i.e., $\sum_{y=0}w_i = \sum_{y=1}w_i$, which tends to improve the convergence of the NN to the optimal classifier. When training a classifier between two hypotheses, this choice simplifies the optimal classifier to
\begin{equation}\label{eq:optimal_classified_eq}
s(x_i) = \frac{p(x_i|\mu_1)}{ p(x_i|\mu_0) +  p(x_i|\mu_1)}.
\end{equation}

\Eqn{\ref{eq:optimal_classified_eq}} can be used to write the probability density ratio of hypotheses $\mu_0$ and $\mu_1$ for a single event $x_i$ as~\cite{GARRIDO199825,Cranmer:2015bka}:

\begin{equation}
r(x_i| \mu_1, \mu_0)=\frac{p(x_i|\mu_1)}{p(x_i|\mu_0)} = \frac{s(x_i)}{1-s(x_i)}.
\end{equation}
The estimate $\hat{r}(x| \mu_1, \mu_0)$ for the ratio $r(x; \mu_1, \mu_0)$ is obtained by substituting the optimal decision function $s(x)$ with its NN estimate $\hat{s}(x)$.
This relation enables the probability density ratio of two values of POIs for individual events to be estimated without the need for dimensionality reduction or histograms. The probability density ratio for the dataset is constructed by taking the product of probability density ratios for individual events to compute the test statistic comparing the two hypotheses $\mu_0$ and $\mu_1$. This trick has, for instance, been used to obtain per-event probability density ratios, in data-driven background models~\cite{HDBS-2019-29} and unfolding~\cite{STDM-2020-17} in the ATLAS experiment.

The task of parameter estimation is a composite hypothesis test, but can be performed by comparing the likelihood for two values of a parameter at a time. While it may appear that parameter estimation would require training a separate classifier for each pair of hypotheses being compared, in practice there are more elegant solutions. A single \textit{parametrized network} may be trained to learn a conditional decision function that varies with the hypothesis under consideration (i.e., the different values of the theory parameter)~\cite{Cranmer:2015bka,Baldi:2016fzo}.
If the parametric dependence of a test statistic can be  expressed analytically in terms of the POIs and a finite number of likelihood ratios estimated from binary classifiers, the need for a network to learn the parametric dependence is eliminated.

\subsection{A simple $\mu$S+B model}

The measurement of the signal strength $\mu$ of a process with no interference with the background processes is defined by the mixture model,
\begin{equation}
p(x_i|\mu)  = \frac{\mu\nu_{\text{S}} \; p(x_i|\text{S}) + \nu_{\text{B}} \; p(x_i|\text{B})}{\mu\nu_{\text{S}} +\nu_{\text{B}}},
\end{equation}
where S represents the signal processes, B the background processes, $\nu_{\text{S}}$ the expected signal yield, and $\nu_{\text{B}}$ the expected background yield. One can train a classifier to estimate a decision function separating signal from background events using balanced class weights,
\begin{equation}
s(x_i) = \frac{p(x_i|\text{S})}{ p(x_i|\text{B}) +  p(x_i|\text{S})},
\label{equation:SimpleCaseClassifier1}
\end{equation}
and then compute the per-event probability density ratio,
\begin{equation}
r(x_i|S,B) = \frac{p(x_i|\text{S})}{ p(x_i|\text{B})} = \frac{s(x_i)}{1-s(x_i)}.
\end{equation}
This can be scaled as required to construct the likelihood ratio,

\begin{equation}
\frac{p(x_i|\mu)}{ p(x_i|\mu=0)}  = \frac{1}{(\mu \nu_{\text{S}} + \nu_{\text{B}})}\frac{\mu  \nu_{\text{S}}\; p(x_i|\text{S}) + \nu_{\text{B}} \; p(x_i|\text{B})}{p(x_i|\text{B})}
= \frac{1}{(\mu \nu_{\text{S}} + \nu_{\text{B}})} \Big( \mu \nu_{\text{S}} \; r(x_i|\text{S},\text{B}) + \nu_{\text{B}} \Big)
,
\end{equation}
where $\nu_{\text{S}}$ and $\nu_{\text{B}}$ are estimated from simulation. The output of a single $\mu$-independent classifier is a \textit{sufficient summary statistic}, meaning that it contains all the information necessary to do hypothesis tests over a range of $\mu$. This is guaranteed by the Neyman-Pearson lemma, which states that the likelihood ratio is the optimal observable when comparing two hypotheses. To use this classifier output directly as an estimate of the probability density ratio, stringent requirements would need to be placed on the quality of this estimate. Alternatively, the output of the classifier can be treated as a high-level observable particularly sensitive to $\mu$. For this reason, it is often used as the final observable in histogram-based signal strength measurements. In such analyses, the likelihood is traditionally computed in each bin of a histogram using a multinomial probability model, where the fraction of events in each bin is obtained from simulation and the observed number of events from data~\cite{cowan1998statistical}. In non-linear problems, no single observable is a sufficient summary statistics. The next section developes a formalism to generalizes the procedure presented above.

\subsection{Search-oriented mixture model}

When the hypotheses being tested can be decomposed into components of a mixture model, the learning task can be factorized into a series of simpler classification tasks~\cite{Cranmer:2015bka}. If the only parameters to be estimated can be written as signal strengths, the individual classifiers no longer need to be parametrized as a function of the POIs, since the relation is explicitly known. This analytical decomposition reduces the burden of validating the test statistic from testing its interpolation performance over the full theory parameter space to validating the performance of a small number of classifiers with fixed hypothesis. If every classifier is well-trained and well-calibrated, then their combination may be expected to remain well-behaved, although this must be explicitly verified.

For the description of a general signal strength measurement at the LHC based on final state that receives contributions from multiple physics processes, the probability density can be described as a mixture model:

\begin{equation}
p(x_i|\mu) = \frac{1}{\nu(\mu)} \sum_J^{C_{\text{proc}}} f_J(\mu) \; \nu_J \; p_J(x_i),
\end{equation}
where index $J$ enumerates the $C_{\text{proc}}$ different physics process, $p_J(x_i)$ is the probability density for the event $i$ corresponding to the process $J$, $\nu_J$ the expected yield for that process with $\mu$ at the Standard Model value, and $\nu(\mu) = \sum_J f_J(\mu)\; \nu_J $. Here $\mu$ can either represent a single theory parameter or a vector of theory parameters, and the formalism accommodates both cases. The functions $f_J(\mu)$ describe how each process scales with the parameters of interest and are determined by the theory model. The mixture model can be rewritten as a function of probability density ratios between the different $C_{\text{proc}}$ processes and a single reference,
\begin{equation}
\label{eq:so_mm}
\frac{p(x_i|\mu)}{p_\text{ref}(\vx_i)}
=    \frac{1}{\nu(\mu)}\sum_J^{C_{\text{proc}}} f_J(\mu)  \; \nu_J \;\frac{p_J(x_i)}{p_\text{ref}(x_i)}.
\end{equation}
The probability density ratios, $p_J(x_i)/p_\text{ref}(x_i)$, can be estimated by using a finite number of $\mu$-independent classifiers. In measurements of signal strengths, the dependence on $\mu$ is completely captured by the coefficients $f_J(\mu)$. If the dependence of $p(x_i|\mu)/p_\text{ref}(x_i)$ on the POIs $\mu$ is not known analytically, a parametrized network can be trained instead~\cite{Cranmer:2015bka}.

As will be shown in \Sect{\ref{sec:testStatistic}}, the reference probability density $p_{\text{ref}}$ is arbitrary. Neither the parameter estimate nor the confidence intervals depend on the choice of $p_{\text{ref}}$. It is just introduced to rewrite the model as a function of probability density ratios. The reference probability density has to satisfy $p_{\text{ref}}(x) > 0$ throughput the phase space of the analysis for the ratio to be well-defined. In order to satisfy this constraint, this paper defines a \textit{search-oriented} reference process, built as a combination of all signal processes,
\begin{equation}
\label{eq:referenceDefnition}
p_\text{ref}(x_i)
=    \frac{1}{\sum_K^{C_\text{signals}}\nu_K}\sum_J^{C_\text{signals}} \nu_J \; p_J(x_i),
\end{equation}
with $C_\text{signals}$ as the number of signal processes. This definition ensures that the denominators in \Eqn{\ref{eq:so_mm}} is larger than zero over the entire signal region of an analysis, which is the region that is sensitive to the signal processes.
Here, $p_\text{ref}$ is defined to be independent of $\mu$, which allows the construction of the final profile likelihood ratio that is independent of $p_\text{ref}$ (see \Sect{\ref{sec:systematics}}). The term $p_\text{ref}$ contributes to a constant offset towards $\log{p(x_i|\mu)}$, which can be ignored in the maximization of the likelihood.

The search-oriented mixture model overcomes issues that may arise in alternative formulations of the reference sample. If probability density ratios $p_J(x)/p_{\text{ref}}(x)$ are used to estimate the likelihood ratio in regions of phase-space with $p_{\text{ref}}(x)\approx 0$, the final estimate will depend on a fine-tuned cancellation of large numbers, which is numerically unstable. A pre-selected signal region for the analysis must be defined that ensures $p_\text{ref}(x_i)>0$ throughout the region. This definition of $p_\text{ref}$ ensures that all signal-sensitive regions of the phase space can remain in the pre-selection region. This choice of $p_\text{ref}$ also aids in the sample-efficient training of the individual classifiers. Finally, it may be convenient to define $p_\text{ref}$ such that it can be represented using simulated samples with only positive weighted events. This simplifies the procedure to construct confidence intervals, which are described in \Sect{\ref{sec:NeymanConstruction}}.

\subsection{Robust estimators with ensembles}
\label{sec:ensembling}
In a traditional analysis where a classifier is employed solely for constructing a sensitive observable and where probability density estimation is performed with a histogram, an imperfect training leads to a suboptimal observable and a slightly less sensitive analysis. However, it does not lead to an ill-behaved test statistic, introduce inaccuracies in the measured confidence intervals, or introduce biases in the maximum likelihood estimate of the POIs. These undesirable behaviors are absent because the likelihood of event counts per bin in a histogram can be computed exactly using the Poisson probability density function. In NSBI, the probability density ratios are instead estimated using NNs, making the high quality of these estimates imperative. Since an individual classifier may not perfectly estimate the decision function $s(x_i)$, a series of steps is described to ensure that the estimator $\hat{s}(x_i)$ is well-behaved (as determined by the diagnostic tests described in \Sect{\ref{sec:diagnostics}}).

One possibility is to calibrate $\hat{s}(x_i)$ using simulated samples~\cite{Cranmer:2015bka}, however, achieving accurate and continuous calibration in practice can be technically challenging. Instead, an ensemble of networks is trained~\cite{10.5555/2381019} on bootstrapped samples of the training data, and their average response used to construct a robust estimate of likelihood ratios. The bootstrapping can be implemented either through resampling or using Poisson perturbations to the event weights that correspond to statistical fluctuations~\cite{ATL-PHYS-PUB-2021-011}. This approach helps account for the variance between individual networks, originating from the random initialization of weights and the finite number of events of the training samples. A similar method has previously been used for neural-network-based data-driven background estimates~\cite{HDBS-2019-29} and unfolding of differential cross-sections~\cite{STDM-2020-17} in ATLAS\@. Examining classifier and ensemble performance across different parts of the observable phase space can guide decisions about NN architecture optimization and data preprocessing. Furthermore, iterative optimization is essential to achieve a high level of accuracy in likelihood-ratio estimation.

Multiple diagnostic tests help determine whether the level of precision desired from the ensembles is achieved, which are discussed in \Sect{\ref{sec:diagnostics}}. Ultimately, the full test statistic must be tested on simulated samples at different values of the POIs to ensure that reliable results with the desired precision are consistently produced over the entire parameter range. Since the ensembles are trained on bootstrapped samples, it is possible to use the spread in their predictions to assess the uncertainty due to the finite training data.

% End of text imported from the .//./sections/NSBI.tex input file

\section{Example use case: off-shell Higgs boson production}
\label{sec:ExampleUseCase}

% The next lines are included from the .//./sections/ExampleUseCase.tex input file
The developed NSBI framework is demonstrated using a subset of the samples originally generated in ATLAS for an off-shell Higgs boson production measurement in the $H\to ZZ\rightarrow 4\ell$ decay channel using a sample of proton--proton collisions at $\sqrt{s}=13\,\TeV$ corresponding to an integrated luminosity of $140\,\text{fb}^{-1}$. The original analysis is described in detail in \Refn{\cite{HIGG-2018-32}}, only the details relevant to NSBI are summarized below. For this demonstration only a subset of the physics processes and systematic uncertainties from the original analysis are considered. The results of a complete implementation of NSBI in the measurement of off-shell Higgs production are given in \Refn{\cite{physicsCompanionPaper}}.

When the quantum interference between signal and background processes is negligible, a single observable that optimally separates signal from background contains all the information necessary to perform optimal hypothesis tests over the full range of signal strength values (see \Eqn{\ref{equation:SimpleCaseClassifier1}}). However, this is no longer true when the interference cannot be ignored, as is the case in the off-shell Higgs boson analysis. Large interference contributions cause the kinematic distributions to change non-linearly with the signal strength $\mu$. \Refn{\cite{Ghosh_jrjc_proc}} demonstrates that the use of NSBI can fully account for these non-linear effects and recover the sensitivity that was lost by the single observable test statistic approach.

The simulated samples used in the study include those for the $gg \rightarrow H^* \rightarrow ZZ \rightarrow 4\ell$ signal-only (S) process, $gg \rightarrow ZZ \rightarrow 4\ell$ background-only (B) process, and the combined simulation including interference effects $gg \rightarrow (H^*\to) \rightarrow ZZ \rightarrow 4\ell$ (SBI$_1$), where the subscript indicates that $\mu=1$ was used for the simulation. These samples, describing the gluon fusion (ggF) production channel, are re-used for the demonstration in this paper.
The full probability model for the ggF production can be expressed as\footnote{In principle, a coupling modifier parameter that scales the signal amplitude is a complex number, and may lead to a phase contributing to the interference term in the cross-section computation. This would require the measurement of two independent parameters of interest. In this analysis, the modifier $\sqrt{\mu}$ is assumed to be a positive real number, and therefore only the inference of one parameter of interest $\mu$ is required.}
\begin{equation}
p_{\text{ggF}}(x |\mu) = \frac{1}{\nu_\text{ggF}(\mu)} \left[(\mu-\sqrt{\mu}) \, \nu_{\text{S}} \, p_{\text{S}} (x) + \sqrt{\mu} \, \nu_{\text{SBI}_1} \, p_{\text{SBI}_1}(x) + (1-\sqrt{\mu})\nu_{\text{B}} \, p_{\text{B}}(x) \right],
\end{equation}

where $\nu_\text{ggF}(\mu) = (\mu-\sqrt{\mu}) \, \nu_{\text{S}} + \sqrt{\mu} \, \nu_{\text{SBI}_1} + (1-\sqrt{\mu})\nu_{\text{B}}$. The contribution from the interference (I) is calculated as $p_{\text{I}} = p_{\text{SBI}_1} - p_{\text{B}} - p_{\text{S}}$, and it is this inference effect that introduces the non-linearity in $\mu$. Following \Refn{\cite{HIGG-2018-32}}, the full probability model $p_{\text{ggF}}(x |\mu)$ and $\nu_\text{ggF}(\mu)$ are functions of $\mu$, while $p_{\text{SBI}_1}(x)$ and $\nu_{\text{SBI}_1}$ are not, and assume $\mu=1$.
For simplicity, the ggF subscripts are suppressed henceforth.
The reference process definition in \Eqn{\ref{eq:referenceDefnition}} leads to  $p_\text{ref}=p_\text{S}$ for this example, and the search-oriented mixture model from \Eqn{\ref{eq:so_mm}} becomes
\begin{equation}
\label{eq:ggmodel}
\frac{p(x |\mu)}{p_\text{S}(x)} = \frac{1}{\nu(\mu)} \left[(\mu-\sqrt{\mu}) \, \nu_S + \sqrt{\mu} \, \nu_{\text{SBI}_1} \, \frac{p_{\text{SBI}_1}(x)}{p_\text{S}(x)} + (1-\sqrt{\mu})\nu_{\text{B}} \, \frac{p_{\text{B}}(x)}{p_\text{S}(x)} \right].
\end{equation}
This density ratio can be constructed from two ensembles, one for each probability density in \Eqn{\ref{eq:ggmodel}}, one to estimate$p_{\text{SBI}_1}(x)/p_\text{S}(x)$ and another one to estimate $p_{\text{B}}(x)/p_\text{S}(x)$. The event selection strategy follows that of Ref.~\cite{HIGG-2018-32} and additionally uses a multivariate discriminant, similar to the discriminant designed in \Refn{\cite{HIGG-2018-32}}, but used here only to define the signal and control regions, i.e., parts of the data without sensitivity to the parameter of interest. Control regions can be used to validate the background model and potentially fit background-related NPs. The remainder of this section will describe the input features and architecture for the networks trained for these tasks, and the systematic uncertainty model used in this example.

\subsection{Input features}

With enough training events, deep NNs can learn from low-level input features such as the four-momenta of all observed final state particles, and can then capture all higher-level correlations. However, in the regime of limited simulated samples, as is often the case at LHC experiments, there is a benefit to using a set of physics-motivated high-level observables that completely describe the observed final state.

The set of observables used to train the network in this demonstration are described in \Tab{\ref{tbl:analZZ4l_ggFNNInputs}}. The Higgs boson decay into Z bosons is described with seven kinematic observables: $\cos\theta^{*}$, $\cos\theta_{1}$, $\cos\theta_{2}$, $\phi_{1}$, $\phi$,  $m_{Z1}$ and $m_{Z2}$. These have traditionally been used to construct a discriminant based on matrix-element calculations, and contain all relevant information to distinguish the Higgs boson signal process from the background~\cite{Bolognesi:2012mm}. Combined with the observables $m_{4\ell}$, $\pt^{4\ell}$ and $\eta_{4\ell}$ for Higgs production, these can be used to calculate the four-momenta of all final-state leptons in the $ZZ\rightarrow 4\ell$ decay channel. Further details on the observables and event selection can be found in \Refn{\cite{HIGG-2018-32}}.
\begin{table}[tbp]
\centering
\caption{List of input variables for the NN. For additional details, see Ref.~\cite{HIGG-2018-32}.}
\begin{tabular}{c|c}\hline\hline
\textbf{Variable} & \textbf{Definition}\\\hline
\multicolumn{2}{@{}l}{Production Kinematics} \\
\hline
$m_{4\ell}$ & Four-lepton invariant mass\\
$\pt^{4\ell}$ & Four-lepton transverse momentum\\
$\eta^{4\ell}$ & Four-lepton pseudo-rapidity\\
\hline
\multicolumn{2}{@{}l}{Decay Kinematics} \\
\hline
$m_{Z_1}$ & $Z_1$ mass \\
$m_{Z_2}$ & $Z_2$ mass\\
$\cos\theta^*$ & cosine of the Higgs boson decay angle\\
$\cos\theta_1$ & cosine of the $Z_1$ boson decay angle \\
$\cos\theta_2$ & cosine of the $Z_2$ boson decay angle \\
$\phi$ & angle between $Z_1, Z_2$ bosons decay planes\\
$\phi_1$ & $Z_1$ decay plane angle\\
\hline\hline
\end{tabular}
\label{tbl:analZZ4l_ggFNNInputs}
\end{table}

\subsection{Network architecture and training}
The classifiers trained in this demonstration are all feed-forward dense networks and comprise five hidden layers with $1\,000$ nodes each, and a swish activation~\cite{DBLP:journals/corr/abs-1710-05941}, followed by an output layer with a single node and a sigmoid activation. The events were split into training and test sets using the $k$-fold method with $k=10$~\cite{3327fbe9-39c0-3916-9425-5d6fb128f17c}, and a bootstrapped sample was generated from the training set to train each network in an ensemble. A weighted binary cross-entropy loss function that accounts for event weights is used to train the networks with the Nadam optimizer~\cite{9727247} in
TensorFlow~\cite{tensorflow_developers_2022_6519082}. The training required large-scale graphics processing unit (GPU) infrastructure~\cite{Megino:2024htf}, consisting of several hundred Nvidia T4 and Nvidia A100 GPUs. The final networks used in this paper, with 500 networks used in an ensemble, required approximately $4\,000$ GPU hours to train. The training of the ML model for the NSBI analysis can be parallelized since independent NNs are used per process and per source of systematic uncertainty~\cite{refId0}. An example of hyperparameter optimization and training strategy used for a full NSBI analysis is provided in \Refn{\cite{physicsCompanionPaper}}.

\subsection{Systematic uncertainties}
\label{sec:example_sys_unc}

At the LHC\@, systematic uncertainties in the modeling of physics processes are often considered in terms of their effect on the shape of distributions and on the expected yields (the overall normalization). This separation can be carried forward to NSBI\@, as it will be shown in \Sect{\ref{sec:nuisance_parameters}}.  Two systematic uncertainties from the original
study in \Refn{\cite{HIGG-2018-32}} are implemented to demonstrate the treatment of NPs that either modify both the shape $p(x|\mu)$ and yield $\nu(\mu)$ of distributions or only the yield $\nu(\mu)$. These are:
\begin{itemize}
\item \textbf{Missing higher-order QCD uncertainty}: the uncertainty in the missing higher-order QCD corrections to the ggF processes in perturbation theory, which modify both the shapes of the kinematic distributions and expected yields.
\item \textbf{Luminosity uncertainty}: the uncertainty in the integrated luminosity measurement of ATLAS\@, affecting only the expected yields.
\end{itemize}
These two uncertainties are used as examples in \Sect{\ref{sec:resultComparison}} to demonstrate how systematics uncertainties can be implemented in an NSBI analysis. They do not constitute a realistic systematic uncertainty model for an off-shell Higgs boson production measurement, which can include over 100 different sources in the $H^*\to ZZ\to 4\ell$ analysis~\cite{HIGG-2018-32}.

% End of text imported from the .//./sections/ExampleUseCase.tex input file

\section{Diagnostics}
\label{sec:diagnostics}

% The next lines are included from the .//./sections/Diagnostics.tex input file
The precise estimate of likelihood ratios is crucial for a robust final result, and therefore the classifiers used in the framework described in \Sect{\ref{sec:NSBI}} require additional scrutiny compared with classifiers used in traditional histogram-based analyses. In addition to traditional visualizations of classifier performance, such as the receiver operating characteristic (ROC) curve and the distribution of the classifier output, this section describes a list of additional diagnostic tools that help in the validation of the likelihood ratio estimate and find areas of potential room for improvement at the level of detail required for NSBI analysis.

\subsection{Reweighting closure}

\begin{figure}[tbp]
\centering
\subfloat[]{\includegraphics[width=0.45\textwidth]{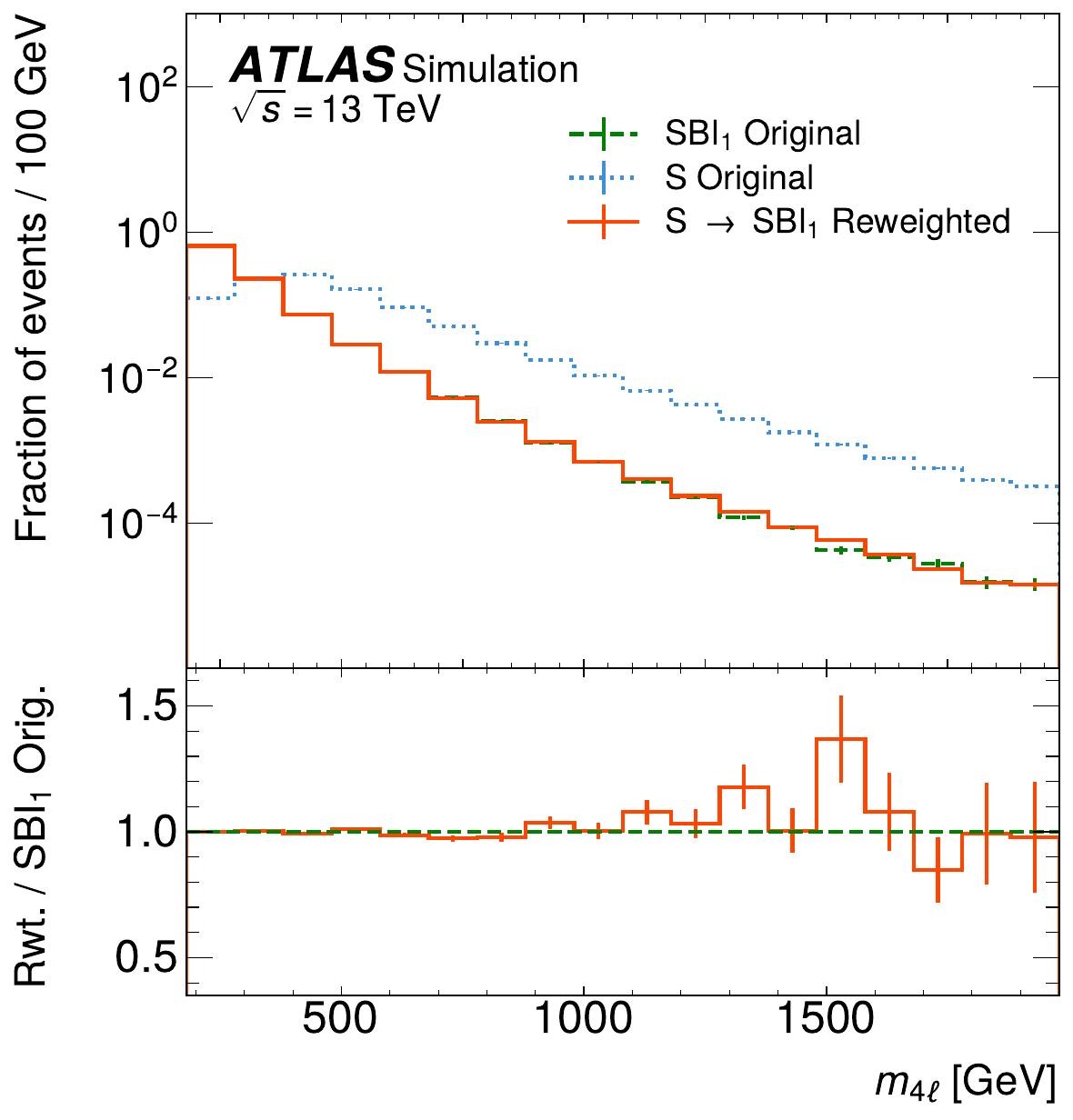}\label{fig:Diagnostics_1D_m4l_sbi}}\hfill
\subfloat[]{\includegraphics[width=0.45\textwidth]{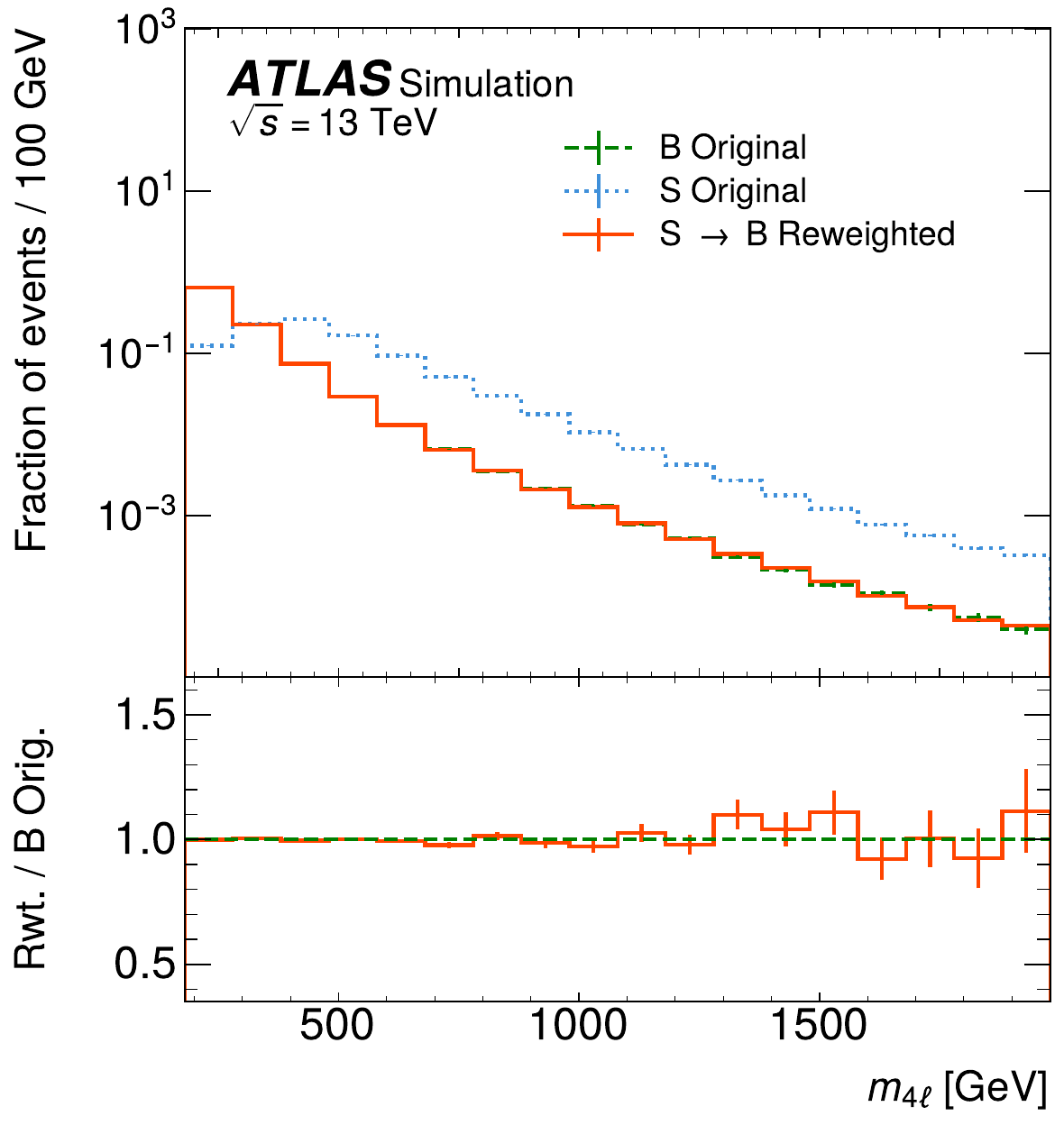}\label{fig:Diagnostics_1D_m4l_b}}
\\
\subfloat[]{\includegraphics[width=0.45\textwidth]{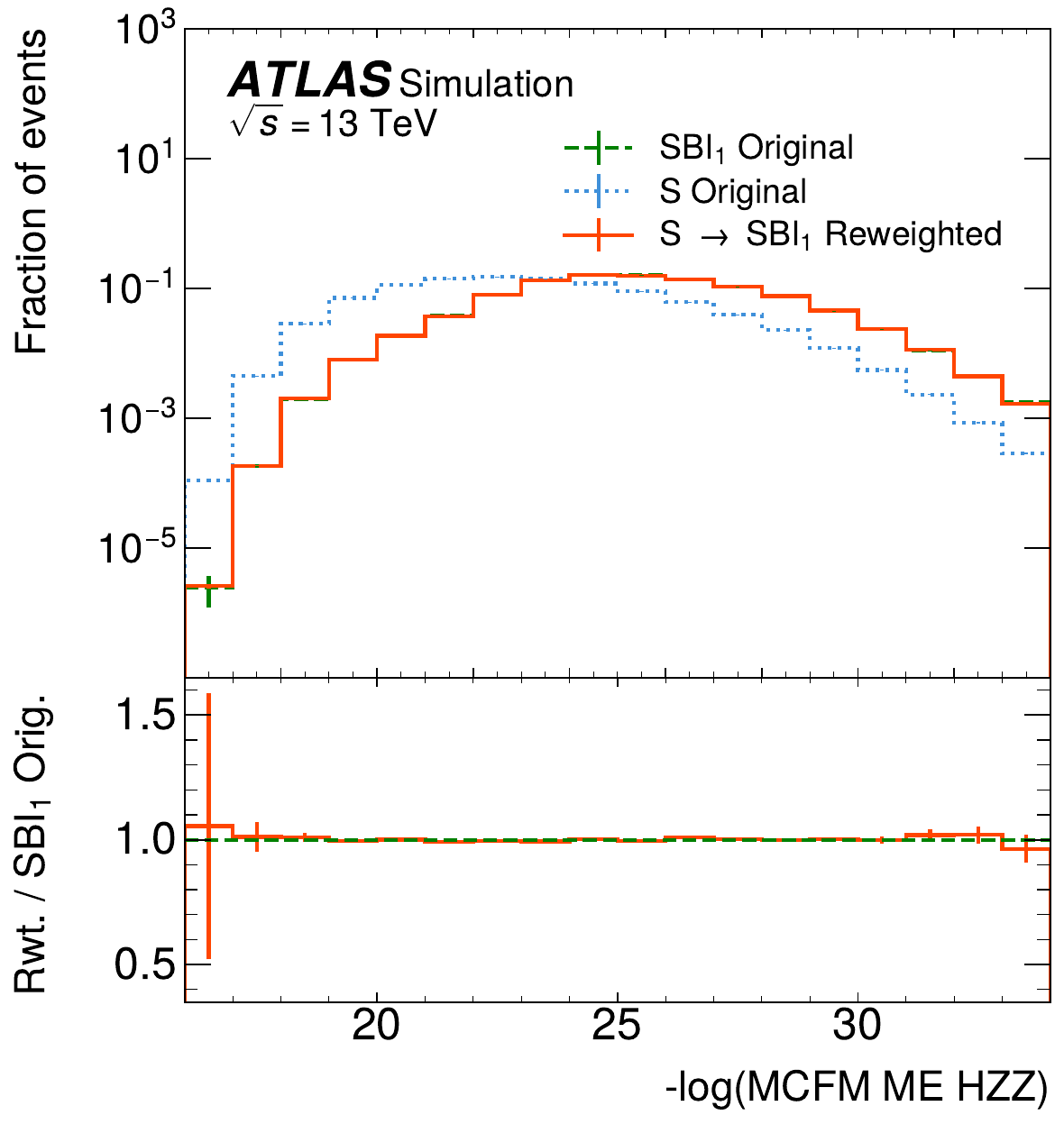}\label{fig:Diagnostics_1D_mcfm_sbi}}\hfill
\subfloat[]{\includegraphics[width=0.45\textwidth]{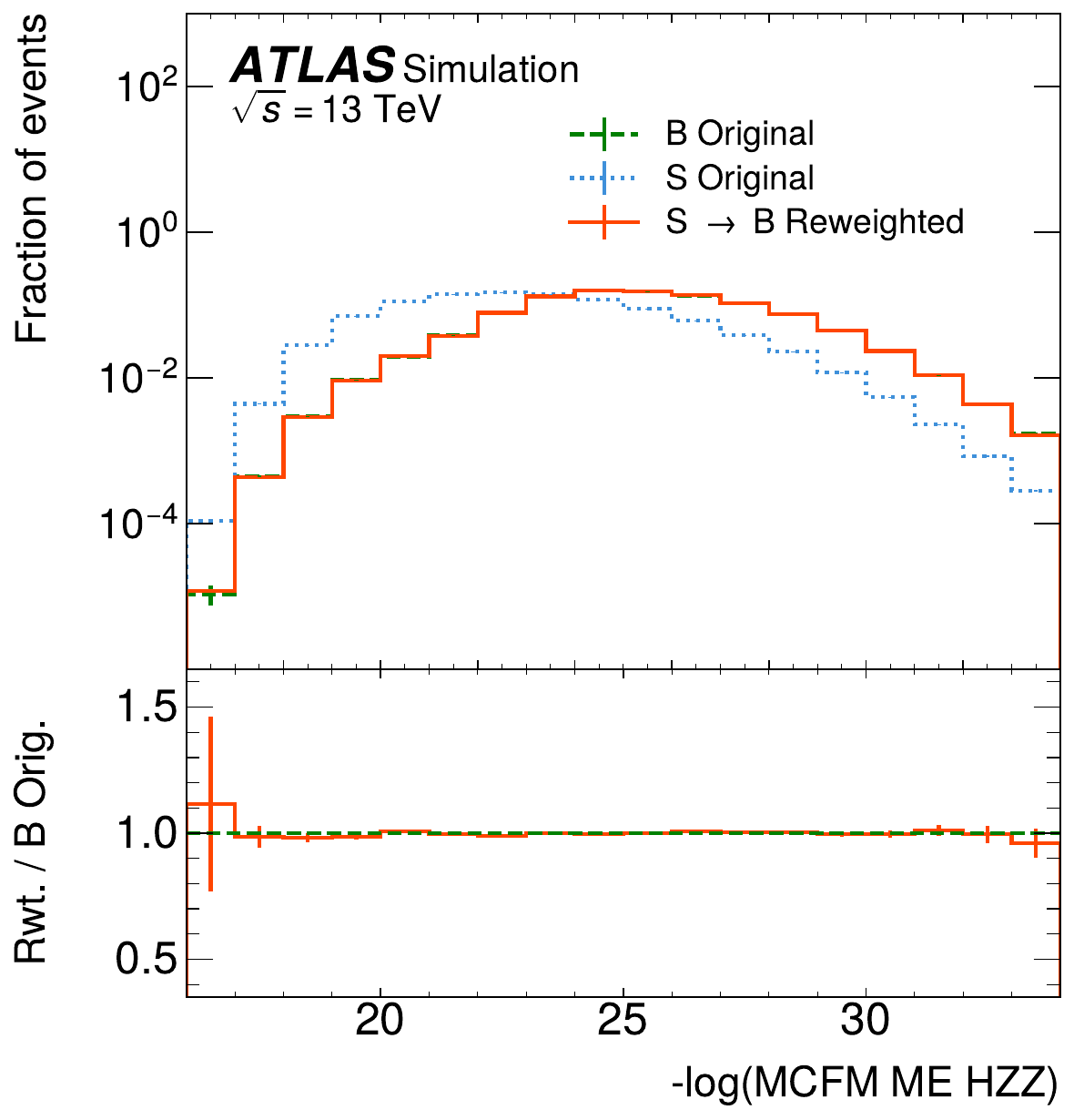}\label{fig:Diagnostics_1D_mcfm_b}}
\caption{One-dimensional reweight closure diagnostic with $m_{4\ell}$ \protect\subref{fig:Diagnostics_1D_m4l_sbi} between the $\text{S}$ and $\text{SBI}_1$ samples and \protect\subref{fig:Diagnostics_1D_m4l_b} between the $\text{S}$ and $\text{B}$ samples. The same diagnostics, using a high-level observable that represents the squared matrix-element for the $gg\rightarrow H \rightarrow ZZ\rightarrow 4l$ process from reconstructed quantities computed using MCFM~\cite{Campbell:1999ah} \protect\subref{fig:Diagnostics_1D_mcfm_sbi} between the $\text{S}$ and $\text{SBI}_1$ samples and \protect\subref{fig:Diagnostics_1D_mcfm_b} between the $\text{S}$ and $\text{B}$ samples. The first diagnostic is an example for an observable directly used in the network training, the second diagnostics is an example of the network's ability to learn high-level physics observables that are not used directly for training. The original reference sample (dotted blue line), is reweighted (solid orange line) using the likelihood ratio estimated with ensembles of NNs to match the target (dashed green line). The lower panels show the ratio of the reweighted reference sample and the original sample. The histograms shown here are examples of a larger validation strategy which requires verifying the closure with many different distributions, possibly in more than one dimension.}
\label{fig:Diagnostics_1D}
\end{figure}

If an ensemble of NNs has estimated the likelihood ratio of two classes $A$ and $B$ accurately, it can be used to reweight samples from one class to another. The normalized distribution of samples from $B$ represented by per-event weights $w_i$, when reweighted as $w_i \; \hat{r}(x_i|A,B)$, where $\hat{r}(x_i|A,B)$ is the NN-based estimate of the likelihood ratio, should result in per-event weights that reproduce the distribution of samples from $A$.

\begin{equation}
\hat{r}(x_i|A,B) \; p(x_i|B) \sim p(x_i|A).
\end{equation}
Any discrepancies indicate a failure of the ensemble of NNs to correctly estimate the likelihood ratio in some part of the phase space. The normalization of the weights $w_i$ restricts the test to differences in the shape of the distribution. Examples of good reweight closure are shown in \Fig{\ref{fig:Diagnostics_1D}} to validate the $p_{\text{SBI}_1}/p_{\text{S}}$ and $p_{\text{B}}/p_{\text{S}}$, using a one-dimensional histogram of the $m_{4\ell}$ observable. Additional comparisons can be made by taking higher-dimensional projections of the full input phase space, but the visualization is challenging for more than two dimensions. The closure is also shown for high-level observables that were not explicitly used in the training, in this case, a matrix-element-based observable that is known to be good summary statistic~\cite{CMS-HIG-18-002}.

An independent classifier, such as a deep NN or a boosted decision tree, can be trained with the goal of separating between events from class $A$ and the reweighted events from $B$ to identify any high-dimensional mismatches between the distributions~\cite{Cranmer:2015bka}. A perfect reweighting would lead to the failure of this independent classifier to reach the goal, indicated by an area under the ROC curve (AUC) of 0.5. Such \textit{classifier tests} provide a multi-dimension probe of the quality of the classifier, albeit in a limited range of the complete phase space~\cite{Krause:2021ilc,Kansal:2022spb}.The classifier test can be performed using larger NNs with bigger capacity than the NNs used to learn the density ratios.

A related tool, the \textit{normalization closure},
\begin{equation}
\sum_{i\in{A}} w_i \frac{p_{{B}}(x_i)}{p_{{A}}(x_i)}=\sum_{i\in{B}} w_i,
\end{equation}
should also be explicitly verified. This simple test can highlight problems if the numerical precision of the training and inference are insufficient to correctly describe events with $s(x_i) \approx 0$ or $s(x_i) \approx 1$.

\subsection{Calibration closure}
\begin{figure}[tbp]
\centering
\subfloat[]{\includegraphics[width=0.47\textwidth]{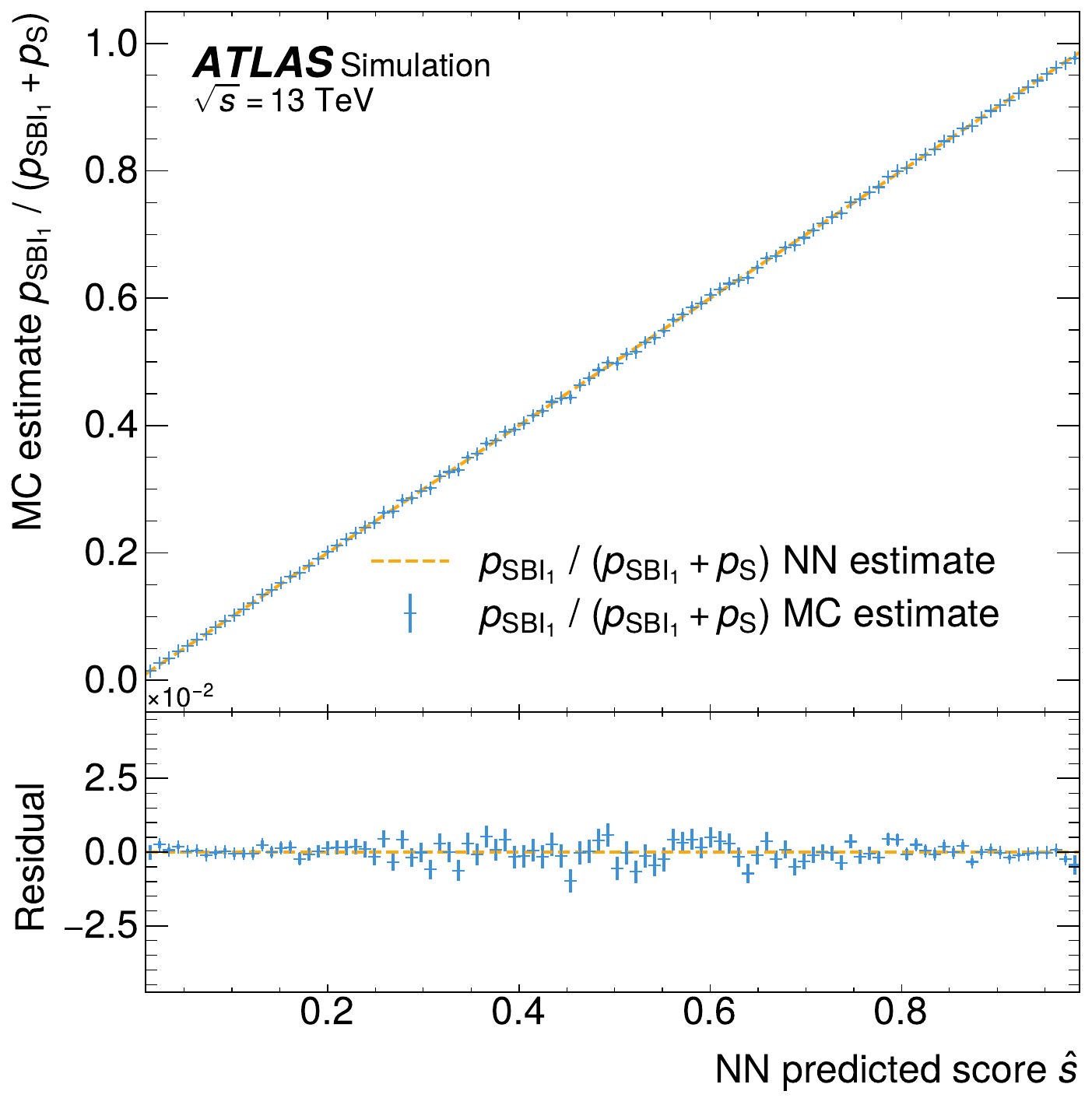}\label{fig:calib_curves_sbi}}\hfill
\subfloat[]{\includegraphics[width=0.47\textwidth]{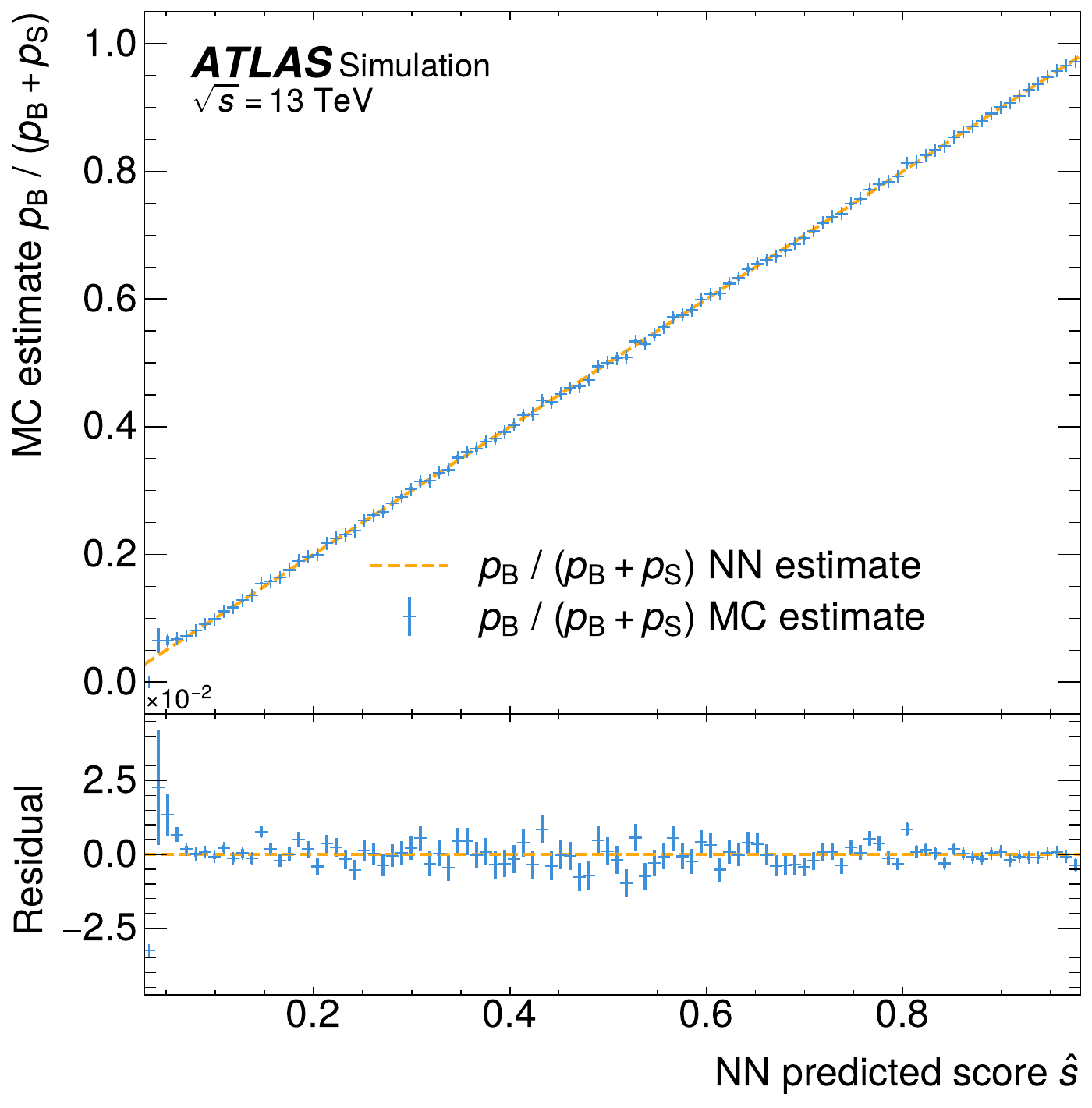}\label{fig:calib_curves_b}}

\caption{Calibration curves comparing ensemble estimated $\hat{s}(x_i)$ with the expected value from binned MC simulated samples, for the validation of the \protect\subref{fig:calib_curves_sbi} $p_{\text{SBI}_1}/\PRE$ and \protect\subref{fig:calib_curves_b} $p_\text{B}/\PRE$ probability density ratio estimations. The residuals, defined as the difference of the MC estimate and the NN estimate, are shown in the bottom panels. The error bars indicate the uncertainty due to the finite number of simulated events in the MC estimate of the density ratio.}
\label{fig:calib_curves}
\end{figure}

Another useful visualization of the NN performance is the calibration curve. If the distribution of predicted relative probability $\hat{s}(x_i)$ from the ensemble of NNs is binned, then the fraction of events in each bin from the first class provides an empirical MC estimate of the mean $s(x_i)$ in that bin. For a well-calibrated classifier, the binned estimate should match the estimate in each bin. \Fig{\ref{fig:calib_curves}} shows the calibration curves for the estimators of $p_{\text{SBI}_1}/\left(p_S+p_{\text{SBI}_1}\right)$ and $p_{\text{B}}/\left(p_S+p_{\text{B}}\right)$ using ensemble predictions. The calibration curves for a well-calibrated classifier is represented as a diagonal line.

\subsection{Spread in ensemble predictions}

\begin{figure}[tbp]
\centering
\subfloat[]{\includegraphics[width=0.45\textwidth]{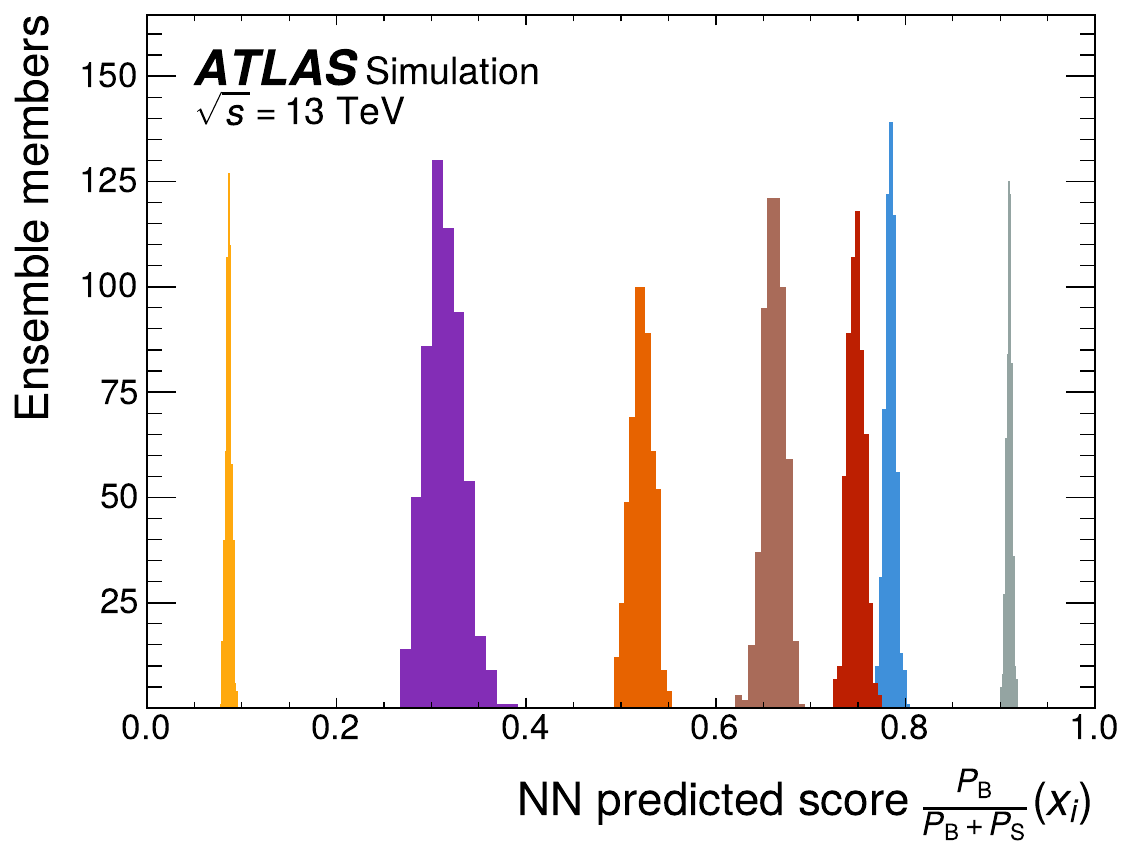}\label{fig:ensemble_pred_scores_b}}\hfill
\subfloat[]{\includegraphics[width=0.45\textwidth]{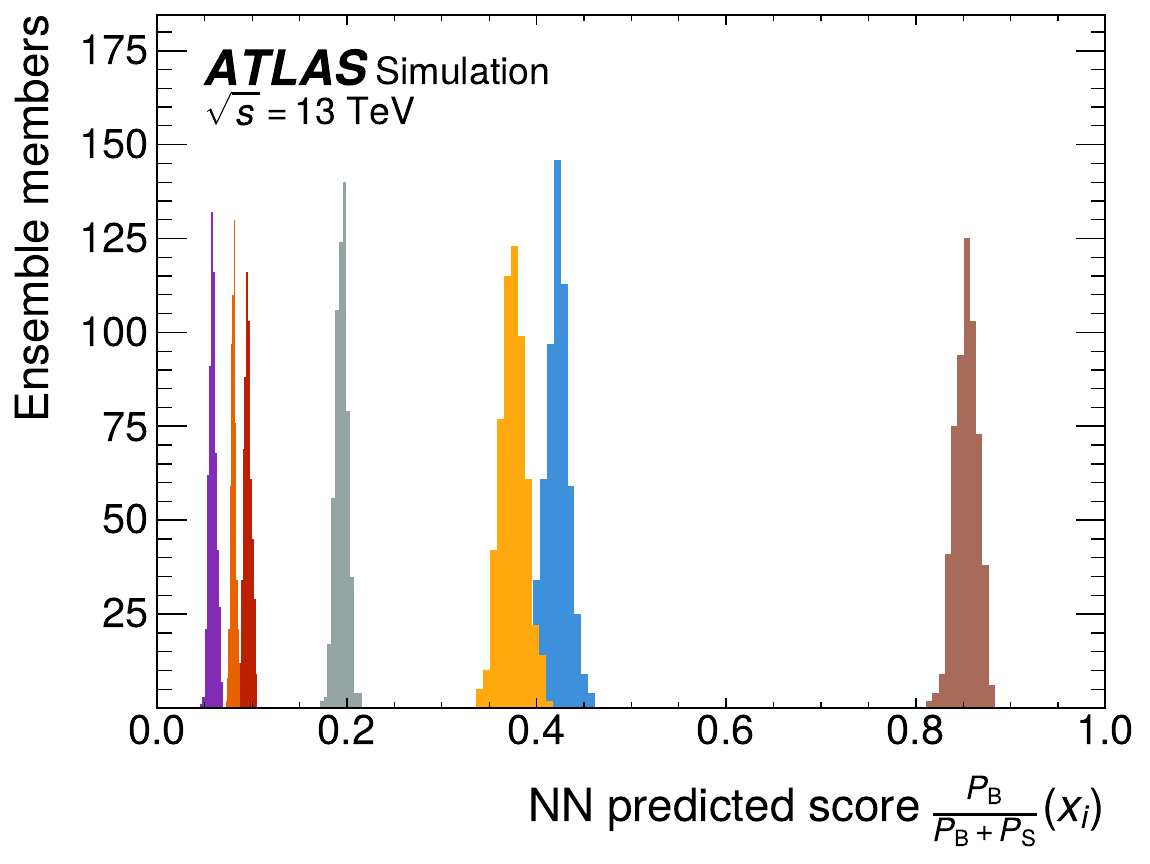}\label{fig:ensemble_pred_scores_s}}

\caption{The distribution of NN output for example events (in different colors), calculated from an ensemble of classifiers trained to separate B from S samples, evaluated on \protect\subref{fig:ensemble_pred_scores_b} seven example events from B and \protect\subref{fig:ensemble_pred_scores_s} seven example events from S. A larger spread indicates a larger uncertainty in the NN score for the event from the ensemble.}
\label{fig:ensemble_pred_scores}
\end{figure}

An ensemble of networks is trained for each classification task, as discussed in Section~\ref{sec:ensembling}. Large spreads in the predictions for the same event reveals regions of phase space for which the number of training events is a limitation, and this can inform the optimization of the training strategy. \Fig{\ref{fig:ensemble_pred_scores}} shows ensemble distributions for a few example events where wider distributions indicate larger uncertainties. The propagation of these uncertainties in the estimated probability density ratios requires careful consideration of their correlated impact on the final parameter estimate, and is described in \Sect{\ref{sec:syst_stat_uncertainties}}.

\subsection{Additional diagnostics}

Additional diagnostic plots may be used to assess the performance of the method, motivated by analysis-specific considerations. In addition to validating the individual estimated probability density ratios $p_J(x_i)/p_\text{ref}(x_i)$ that form the mixture model, the combined probability density ratio $p(x_i | \mu)/p_\text{ref}(x_i)$ can also be validated using the discussed diagnostic tools. The inference can in addition be validated on independent samples simulated at values of $\mu$ that were not used for training. The probability ratio estimates should also be validated in data using control regions. The final performance of the analysis method can also be verified on simulated datasets, sampled over a wide range of values of the parameter of interest to ensure that the correct maximum likelihood estimate is consistently obtained, and this validation is shown in the companion physics analysis paper~\cite{physicsCompanionPaper}.

As a further cross-check, the analysis method can be tested on samples simulated with a different event generator, on samples simulated with a shifted value of a NP, or with different injected POI values. This method provides additional interpretable per-event quantities to examine, i.e., the estimated probability density ratios between different theory hypotheses for a given event. These quantities can be studied as functions of several observables to understand the sub-category of events that influence the overall test statistic in favor of one hypothesis over another. A few examples are discussed in \Sect{\ref{sec:resultComparison}}.

% End of text imported from the .//./sections/Diagnostics.tex input file

\section{Systematic uncertainties}
\label{sec:systematics}

% The next lines are included from the .//./sections/Systematics.tex input file
A major challenge in applying NSBI to LHC data is addressing systematic uncertainties. Each individual source of systematic uncertainty is represented by a NP $\alpha_k$. Collectively the NPs are represented by a vector $\alpha$. The NPs may be constrained by an auxiliary measurement which provides a nominal value $a_k$ and an externally provided uncertainty $\delta_k$ on the nominal value.

In principle, classifiers can be conditioned on NPs in an analysis to propagate uncertainties through to the final inference step~\cite{Cranmer:2015bka,Ghosh:2021roe}. In practice, this is not feasible for all NPs in an analysis. First, due to computational costs, as samples are typically generated by varying a single NP at a time, with no training samples available where multiple parameters vary simultaneously.  Second, because only three sets of samples are typically available per NP: one at the nominal value $\alpha_k^
{(0)} = a_k$ and the others at variations $\alpha_k^{(-)} = a_k - \delta_k$ and $\alpha_k^{(+)} = a_k + \delta_k$. These sets are insufficient for a network to learn the full parametric dependence. Finally, validating the interpolation capabilities of a classifier across all regions of this high-dimensional space of NPs would be challenging even if the classifier were parametrized in all of them.

Instead, this paper extends the systematic uncertainty framework already in place for histogram-based analyses to an unbinned multi-dimensional setting, and incorporates it into NSBI\@. While in a histogram-based analysis the impact of a NP is estimated per bin, for NSBI it is estimated per event, and the interpolation between NP values is also performed using traditional methods, rather than relying on the networks to learn it. Moreover, the impact of systematic uncertainties from independent sources is treated independently, following the standard practice at the LHC.

\subsection{Nuisance parameters in the likelihood ratio function}
\label{sec:nuisance_parameters}
In a histogram-based analysis at the LHC, the impact of systematic uncertainties is typically propagated into the likelihood using \textit{vertical interpolation}~\cite{Cranmer:2014lly}. The impacts of different NPs on the measured yields are considered to be independent of each other and of the POIs,
\begin{equation}
\nu_J(\alpha) = \nu_J(\alpha^{(0)})\prod_k^{N_\text{syst}} G_J(\alpha_k),
\end{equation}
for $N_\text{syst}$ NPs with $G_J(\alpha_k) = \nu_J(\alpha_k)/\nu_J(\alpha_k^{(0)})$.
The functions $G_J(\alpha_k)$ are chosen to smoothly interpolate between their three known values at the points $\alpha_k^{(-)}$, $\alpha_k^{(0)}$ and $\alpha_k^{(+)}$, which are determined from the available simulations~\cite{Cranmer:2014lly}. The choice of a differentiable interpolation function facilitates the computation of pulls and impacts, detailed in \Sect{\ref{sec:pulls}}.

Extending this formalism and the corresponding assumptions to a per-event approach, \Eqn{\ref{eq:so_mm}} can be updated to incorporate NPs as
\begin{equation}
\label{eqn:factorisedGeneralForm}
\begin{split}
\frac{p(x_i|\mu,\alpha)}{\PRE(x_i)}
= \frac{1}{\nu(\mu,\alpha)}\sum_J^{C_{\text{proc}}} f_{J}(\mu) \; \nu_J \; \frac{p_J(x_i)}{\PRE(x_i)} \; \prod_k^{N_\text{syst}} G_J(\va_k) \; g_J(x_i, \va_k),
\end{split}
\end{equation}
with $\nu(\mu,\alpha)=\sum_J^{C_{\text{proc}}} f_{J}(\mu) \; \nu_J \;\prod_k^{N_\text{syst}} G_J(\va_k)$. The contribution to the per-event probability density ratio from each NP comes from $g_J(x_i, \va_k) = p_J(\vx_i,\va_k )/p_J(x_i)$,
where $p_J(x_i)$, $f_J(\mu)$, $\nu_J$ and $p_\text{ref}(x_i)$ are defined at $\alpha^{(0)}$.

As with the functions $G_J(\alpha_k)$, the functions $g_J(x_i, \va_k)$  are chosen
to interpolate between the three known values at $\alpha_k^{(0)}$ and for the alternate cases $\alpha_k^{(\pm)}$ for each event, using the same interpolation strategy. For the nominal case $g_{J}(x_i,\alpha_k^{(0)}) = 1$, and for $\alpha_k^{(\pm)}$, the probability density ratios $g_{J} (x_i, \va_k^{(\pm)})$ are estimated per event by training ensembles of classifiers. These classifiers are trained to separate nominal samples $p_J(x_i)$ from systematic variation samples $p_J(\vx,\va_k^{(\pm)} )$, with one ensemble per physics process, per NP, and per variation.

Once the functions $g_J (x_i, \va_k^{(\pm)})$ are determined, these can even be used to replace the alternative simulations altogether in an analysis. This method was used to describe modeling uncertainties in the previous ATLAS analysis of the off-shell Higgs boson production~\cite{HIGG-2018-32} and more recently in the measurement of the $WH$ and $ZH$ production with Higgs boson decays into bottom and charm quarks~\cite{HIGG-2020-20}. The diagnostic tests described in \Sect{\ref{sec:diagnostics}} are also useful tools to validate these NNs, although they can be less illuminating if the systematic variation is very small (leading to $s(x_i) \approx 0.5$ for all events).

NSBI not only constructs a more sensitive analysis in the entire phase space of the POIs $\mu$, but also in the space of the NPs $\alpha$~\cite{Ghosh:2021roe}. As with histogram analyses, it is important to ensure that an NSBI analysis does not overconstrain a NP, i.e., it does not constrain the uncertainty on NP beyond the externally prescribed uncertainty. Overconstraining might indicate that the modeling of the systematic uncertainty is oversimplified or the fit is exploiting aspects of the systematic uncertainty model that are not known well, for instance in the case of two-point theory uncertainties~\cite{Ghosh:2021hrh}. An analysis of the pulls on the NPs and impacts, described further in \Sect{\ref{sec:pulls}}, and the use of alternative modeling of the systematic uncertainties, such as splitting the NP into independent sub-components, can reveal such issues. Furthermore, LHC experiments often quantify the uncertainties in the systematic uncertainties themselves, and on models of correlation between different components of systematic uncertainties~\cite{PERF-2014-02,STDM-2015-01}. Such challenges are often discussed in the context of model misspecification in ML literature.

The parametrization presented here generalizes the model used in traditional histogram-based analyses by introducing NNs to describe the shape variation of each source of systematic uncertainties. Precision measurements can have a large number of systematic uncertainties. In these cases, the full NSBI model will have a large number of NNs, which need to be evaluated for each event when evaluating the log-likelihood ratio. The practical use of a NSBI analysis requires dedicated computational and software infrastructure.

\subsection{The profile log-likelihood ratio}
\label{sec:testStatistic}
The full test statistic, based on a profile log-likelihood ratio~\cite{Cowan:2010js}, can be constructed from \Eqn{\ref{eqn:factorisedGeneralForm}} by considering all events in the observed data, adding a Poisson term modeling the total rate and including Gaussian constraint factors for the NPs. If $N_\text{data}$ is the number of events in observed data $\mathcal{D}$,

\begin{equation}\label{eqn:Lwithconstraint}
\begin{split}
\frac{L_{\text{full}}(\mu,\alpha|\mathcal{D})}{L_\text{ref}(\mathcal{D})}=\text{Pois}&(N_\text{data}|\nu(\mu,\alpha))
\;
\prod_i^{N_\text{data}}\frac{p(x_i|\mu,\alpha)}{\PRE(x_i)}
\;
\prod_k\text{Gaus}(a_k|\alpha_k,\delta_k),
\end{split}
\end{equation}
where $a_k$ and $\delta_k$ are the values of the auxiliary measurements and their associated uncertainty, which are used to constrain the source of systematic uncertainty associated with the NP $\alpha_k$. The likelihood corresponding to the reference probability distribution can be written as $L_\text{ref}(\mathcal{D}) = \prod_i^{N_\text{data}} \PRE(x_i)$. If a NP is unconstrained, the corresponding constraint factor is suppressed. An important class of unconstrained NPs is data-driven normalization parameters.

The profiling step involves an unconditional and a conditional maximum likelihood estimation of \Eqn{\ref{eqn:Lwithconstraint}},
\begin{equation}
\begin{split}
(\widehat{\mu},\widehat{\alpha})&=\underset{\mu,\alpha}{\text{argmax}} \dfrac{L_{\text{full}}(\mu,\alpha)}{L_{\text{ref}}},\\
\widehat{\widehat{\alpha}}(\mu)&=\underset{\alpha}{\text{argmax}} \dfrac{L_{\text{full}}(\mu,\alpha)}{ L_{\text{ref}}},
\end{split}
\end{equation}
where the dependence of $L_{\text{full}}$ on the data $\mathcal{D}$ is implicit, and $L_\text{ref}$ is a constant that does not depend on $\mu$ or $\alpha$. Hence, it does not affect the position of the maxima. The test statistic is constructed by taking the ratio of \Eqn{\ref{eqn:Lwithconstraint}} at these two points in the parameter space. The dependency on $L_{\text{ref}}$ cancels out and the traditional profile log-likelihood ratio is recovered,
\begin{equation}\label{eq:full_test_stat}
\begin{split}
t_\mu &= -2\ln \left(\frac{L_{\text{full}}(\mu,\widehat{\widehat{\alpha}}(\mu))}{L_{\text{full}}(\widehat{\mu},\widehat{\alpha})}\right).
\end{split}
\end{equation}
The use of likelihood ratios instead of likelihoods does not prevent the combination of NSBI and histogram-based analyses, which can be written as
\begin{equation}
\frac{L_\text{comb}(\mu,\alpha)}{L_\text{ref}} = \frac{L_\text{full}(\mu,\alpha)}{L_\text{ref}} \; L_\text{hist}(\mu,\alpha).
\end{equation}
The corresponding test statistic is again independent of $L_\text{ref}$ since it cancels in the profile likelihood ratio~\cite{physicsCompanionPaper}.

\subsection{Effects from the limited size of simulated samples}
\label{sec:syst_stat_uncertainties}

When likelihood ratios are estimated with NNs, an uncertainty may be introduced to account not only for the limited number of simulated training samples, but also for the stochastic nature of the training algorithm. Training ensembles on bootstrapped versions of the training data, as described in \Sect{\ref{sec:ensembling}} provides a natural way to describe both of these effects.

Since the estimator for the density ratio is computed as the mean\footnote{The median, known to be robust to outliers, could also be used.} prediction from an ensemble of networks, the variance of that mean can be estimated by using the bootstrapping technique. The mean of each bootstrapped ensemble is used to estimate a best-fit value of the parameter(s) of interest $\hat{\mu}$, and the variance of these estimates determines the variation of the mean $\Delta \hat{\mu}$ due the finite number of events in the training sample. The variance can be determined at different values of $\mu$ using different Asimov datasets.\footnote{An Asimov dataset is one for which the application of any unbiased estimator for all parameters will provide the true values~\cite{Cowan:2010js}. In unbinned analyses, an approximation of such a dataset can be constructed using a large number of simulated events with appropriate event weights.} Such datasets, at any value of the POIs, can often be constructed from a set of simulations at few \textit{basis points} in this parameter space, using various morphing techniques~\cite{ATL-PHYS-PUB-2015-047,Brehmer:2019xox}. The estimated $\Delta\hat{\mu}$ is an uncertainty in the modeling of the expected probability density of the physics processes, and therefore, it can be introduced as a systematic uncertainty following the spurious signal approach~\cite{ATL-PHYS-PUB-2020-028} frequently employed in unbinned LHC analyses. A NP $\alpha_\text{stat}$ with a Gaussian constraint term is introduced in \Eqn{\ref{eqn:factorisedGeneralForm}} with the modification

\begin{equation}
f_J(\mu) \rightarrow f_J(\mu + \alpha_\text{stat}\, \Delta\hat{\mu}(\mu)  ).
\end{equation}

\subsection{Calculation of pulls and impacts}
\label{sec:pulls}

While the unbinned nature of NSBI poses computational challenges to traditional statistical tools for evaluating and analysing the profile likelihood ratio, this framework enables the direct application of modern computational tools that simplify calculations. The full likelihood ratio (\Eqn{\ref{eqn:Lwithconstraint}}) and the test statistic (\Eqn{\ref{eq:full_test_stat}}) are differentiable functions. Their dependence on the POIs $\mu$ and NPs $\alpha$ is introduced through differentiable functions, either through smooth functions or through NNs that are themselves differentiable.
It is natural to leverage auto-differentiation techniques~\cite{autodiff} to calculate the Hessian matrix of $L_{\text{full}}(\mu,\alpha)$. Auto-differentiation tools can also be leveraged to perform profiling assisted by exact gradients.

Local estimates of NP pulls and impacts rely on the calculation of the Jacobian and Hessian matrices. The local estimate of pulls and impacts is known as the \textit{HESSE} procedure~\cite{James:2296388,RooFit,moneta2011roostatsproject} in high-energy physics. The covariance estimates it provides are a useful metric of the uncertainty during the development of the analysis. The reported uncertainties of the final analysis are calculated with the so-called \textit{MINOS} procedure, which are based on profile likelihood ratio intervals in the asymptotic approximation. In this section, procedures for the local HESSE approximation based on the exact Jacobian and Hessian matrices are derived. The construction of confidence intervals will be described in \Sect{\ref{sec:NeymanConstruction}}.

The HESSE covariance matrix is estimated using the inverse of the Hessian matrix at the maximum likelihood estimate $(\widehat{\mu},\widehat{\alpha})$,

\begin{equation}\label{eq:cov_matrix}
V_{nm} = \left[\frac{1}{2}\frac{\partial^2 \lambda}{\partial{\alpha}_n\partial{\alpha}_m}(\widehat{\mu},\widehat{\alpha})\right]^{-1},
\end{equation}
where $\lambda(\mu,\alpha)= -2\ln (L_{\text{full}}(\mu,\alpha)/L_{\text{ref}})$ and the POI is identified by the index $m=0$ to simplify the notation. The calculation of the Jacobian and Hessian matrices can be parallelized on computing clusters~\cite{refId0}. The pull of the NP $\alpha$ is defined as

\begin{equation}
\frac{\widehat{\alpha}_k-\alpha_{k}^{(0)}}{\sqrt{V_{kk}}}.
\end{equation}

The impact of a systematic uncertainty is an estimate of the propagation of the uncertainty to the POI $\mu$. Like in case of estimation of pulls, the impact of a systematic uncertainty can be approximated by a local expression based solely on the Jacobian and Hessian matrices or by a more accurate, but computationally more expensive, expression based on the maximum likelihood estimate calculated at different parameter values.

The impact of a systematic uncertainty is traditionally estimated by the difference $\Gamma_k^{\text{NP},(\pm)} = \dhatmu(\widehat{\alpha}_k\pm\sqrt{V_{kk}})-\widehat{\mu}$ where $\dhatmu(\widehat{\alpha}_k\pm\sqrt{V_{kk}})$ is the conditional maximum likelihood estimate obtained by keeping the NP associated with the systematic uncertainty fixed at $\alpha_k = \widehat{\alpha}_k\pm\sqrt{V_{kk}}$. Recently~\cite{Pinto:2023yob}, an alternative definition has been proposed where the impact is estimated with the difference $\Gamma_k^{\text{AO},(\pm)} = \widehat{\mu}(a_k\pm\delta_k)-\widehat{\mu}(a_k)$ of the maximum likelihood estimates when the value of the auxiliary observable (AO) $a_k$ representing the auxiliary measurement is shifted by the externally provided uncertainty and when it is kept at its nominal value. The latter definition has been shown to allow for a consistent decomposition of systematic and statistical uncertainties.

Both definitions of the impact of a systematic uncertainty have local definitions which can be calculated efficiently with auto-differentiation. In the case of a single POI, a local estimate of the impact $\Gamma_k^{\text{NP}}$ is given by

\begin{equation}\label{eq:local_impact}
\begin{split}
\Gamma_k^{\text{NP}}&=\frac{\partial\widehat{\mu}}{\partial\alpha_k}\times \sqrt{V_{kk}}=\left[\frac{\partial^2 \lambda}{\partial^2\mu}(\widehat{\mu},\widehat{\alpha})\right]^{-1}\frac{\partial^2\lambda}{\partial\mu\partial\alpha_k}(\widehat{\mu},\widehat{\alpha})\times\sqrt{V_{kk}},
\end{split}
\end{equation}
considerably simplifying the analysis of the profile likelihood ratio. The effect of pulls on the impact can be evaluated by calculating the so-called \textit{pre-fit} impact, which is obtained by replacing $\sqrt{V_{kk}}\to \delta_k$ in \Eqn{\ref{eq:local_impact}}. A local estimate for the alternative definition of impact is given by $\Gamma^{\text{AO}}_k = V_{0k}(\widehat{\mu},\widehat{\alpha})$~\cite{Pinto:2023yob}, where the covariance matrix is defined in \Eqn{\ref{eq:cov_matrix}}. In non-linear likelihoods with  multiple local minima, the local definition also avoids ambiguities that exist when estimating impacts based on finite differences. Further details about these calculations for NSBI using auto-differentiation techniques are described in Ref.~\cite{refId0}.

Performing the maximization of the likelihood ratio in unbinned analyses can be computationally challenging when the number of selected events is large. In NSBI analyses, an additional challenge stems from the need to perform inference of many NNs to calculate the value of the likelihood ratio for each value of the POIs and NPs.

In order to reduce the computational burden, the values of the NNs can be pre-calculated in a trade-off between processing time and memory usage. The time to perform a single maximization in a modern CPU core and the memory used vary between $\mathcal{O}(10)$ minutes and $\mathcal{O}(1)$ gigabytes (GB) for regions with a few thousand events, and $\mathcal{O}(10)$ hours, and $\mathcal{O}(100)\,\text{GBs}$ for regions with few million events. An alternative workflow where the NNs are evaluated during each maximization step could be advantageous in high-performance computing systems where several GPUs can be used in parallel for a single process. In such workflows, memory usage would scale only with the number of features, and not with the number of events.

% End of text imported from the .//./sections/Systematics.tex input file

\section{Neyman construction}
\label{sec:NeymanConstruction}

% The next lines are included from the .//./sections/NeymanConstruction.tex input file
In frequentist statistics, a confidence interval derived from a measurement is expected to cover the true value with a specified probability (e.g., in 68\% or 95\% of experiments). The procedure for building such confidence intervals, referred to as the \textit{Neyman construction}, involves the inversion of the hypothesis tests. When the distribution of the test statistics is not known, e.g. outside the asymptotic regime~\cite{Cowan:2010js}, or for non-linear problems, this distribution can be estimated with the help of a large number of pseudo-experiments generated based on simulated samples~\cite{1937RSPTA.236..333N}.

In the case of NSBI, any residual bias in the estimated probability density ratios may produce a test statistic that does not follow a $\chi^2$ distribution, making this procedure all the more crucial. The procedure for producing such pseudo-experiments, often referred to as \textit{throwing toys}, is well established for histogram-based analyses, where the probability density can be sampled as individual Poisson distributions in each bin. This approach can be extended to an unbinned, multi-dimensional NSBI analysis. The example of the off-shell Higgs production analysis described in \Sect{\ref{sec:ExampleUseCase}} requires a Neyman construction because of the double local minima created by the quantum interference and because of the bias created by the $\mu>0$ condition. These conditions are captured by the simplified analysis presented in this paper and will serve as an example of the produced developed in this section.

\subsection{Generating pseudo-experiments}

Similar to events measured in an actual experiment, pseudo-experiments consist of unweighted events. These can be generated by sampling simulated events with replacement, with the probability of sampling an event determined by its original weight in the Asimov dataset, $w_i^{\text{Asimov}}$. Since the same simulated event can be chosen multiple times in a pseudo-experiment, this count can be represented by a new integer event weight, $w_i^{\text{toy}}$.\footnote{While the integer weights $w_i^{\text{toy}}$ are used for convenience, the constructed pseudo-experiment still behaves effectively like an unweighted dataset.} For a computationally efficient generation of these pseudo-experiments, each simulated event is assigned a $w_i^{\text{toy}}$ sampled from a Poisson random number generator with a mean corresponding to the Asimov weight of the event,
\begin{equation}
w_i^{\text{Asimov}} \rightarrow w_i^{\text{toy}} = \text{Poisson}(w_i^{\text{Asimov}}).
\end{equation}
The generated weights $w_i^{\text{toy}}$ are integers by construction. Since $w_i^{\text{Asimov}}$ represent fractional weights, with a magnitude of $\mathcal{O}(10^{-3})$ for the example described in \Sect{\ref{sec:ExampleUseCase}}, most event are assigned a weight of zero, and a smaller subset is assigned integer weights. A very small fraction of events may be represented multiple times in a single pseudo-experiment ($w_i^{\text{toy}} \geq 2$), similar to the scenario of generating samples via bootstrapping. To generate such pseudo-experiments from a simulated sample, the original number of simulated events must be much larger than the number of events in an individual pseudo-experiment.

\subsection{Overcoming negative weights}
\label{subsec:NC_neg_weights}

The prescription for generating unweighted pseudo-experiments requires the original weights of the simulated events to be non-negative, $w_i^{\text{Asimov}} \geq 0$, since the Poisson distribution is only defined for non-negative values. When the MC simulation sample at a given value of the POIs includes events with negative weights, an alternate sample may be used that consists only of positive weights and covers the support of the original sample. The alternate sample, henceforth referred to as the \textit{reweight reference} sample, will have to first be reweighted to the desired value of the POIs. The samples corresponding to the reference defined in \Sect{\ref{sec:NSBI}} may be a convenient choice for the reweight reference sample because it already covers the entire preselection region and can be defined to comprise only positive-weighted events. Since the reference sample does not need to correspond to a physical process, a very large sample can be simulated at leading-order in perturbation theory and without negative weights. A large reference sample is not only ideal for the network training but also to allow the generation of large number of pseudo-experiments following the methods described here. The reweight reference can be reweighted using \Eqn{\ref{eq:so_mm}} to the desired value of the theory parameter

\begin{equation}
w_i^{\text{rwt-ref}} \rightarrow w_i^{\text{Asimov}} (\mu) = \frac{\nu(\mu)}{\nu_\text{rwt-ref}} \; \frac{p(x_i|\mu)}{p_{\text{rwt-ref}}(x_i)} \; w_i^{\text{rwt-ref}},
\end{equation}
where $p_{\text{rwt-ref}}(x_i)$ is the probability density and $\nu_\text{rwt-ref}$ is the rate for the reweight-reference sample. The probability density ratio $p(x_i|\mu)/p_{\text{rwt-ref}}(x_i)$ can be obtained from ensembles specifically trained for the reweighting procedure, following the same prescription as for the networks used for inference. The estimates can be validated using the same diagnostics described in \Sect{\ref{sec:diagnostics}}, and the new samples are thereby verified to have the same asymptotic properties as the original MC simulation samples. There are also other methods that could be explored to handle negative weighted events~\cite{Backes:2020vka,Nachman:2020fff,Drnevich:2024vfj}.

\subsection{Confidence intervals}

\begin{figure}[tbp]
\vspace{0.1in}
\centering
\subfloat[]{\includegraphics[width=0.49\textwidth]{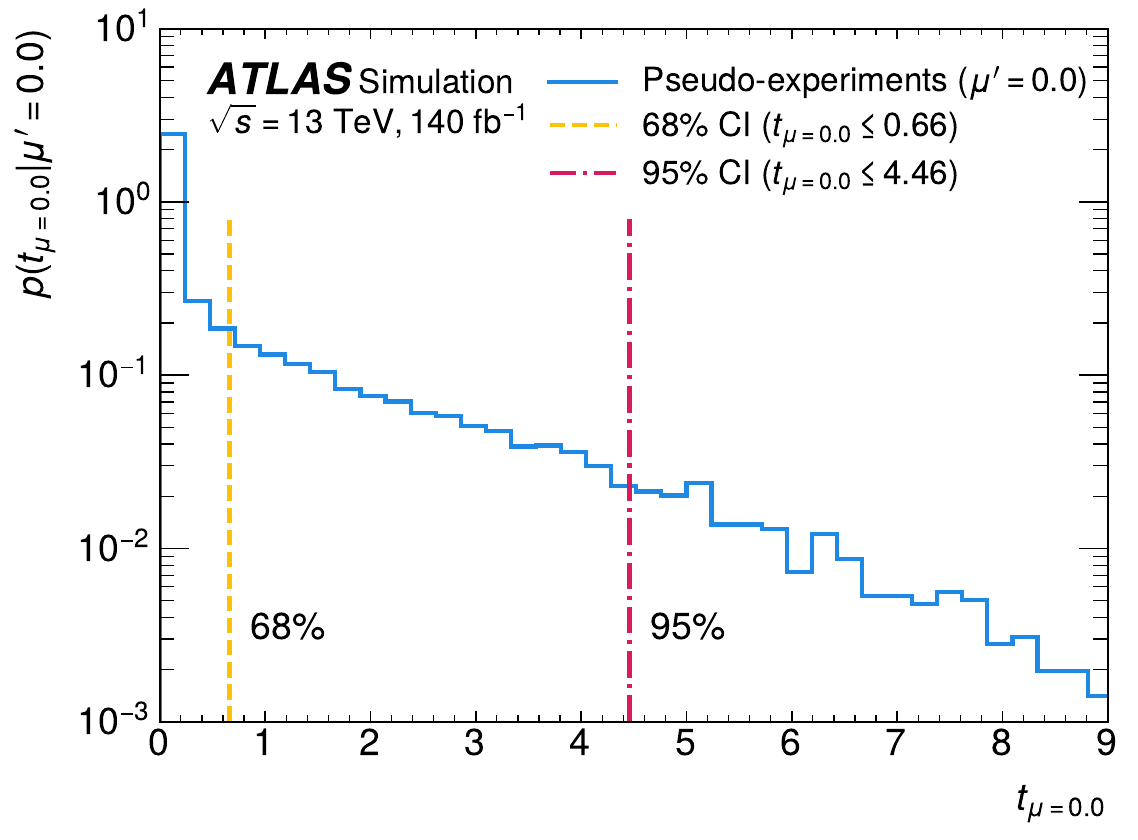}\label{fig:NeymanConstruction_t_mu_1p00_0}}\hfill
\subfloat[]{\includegraphics[width=0.49\textwidth]{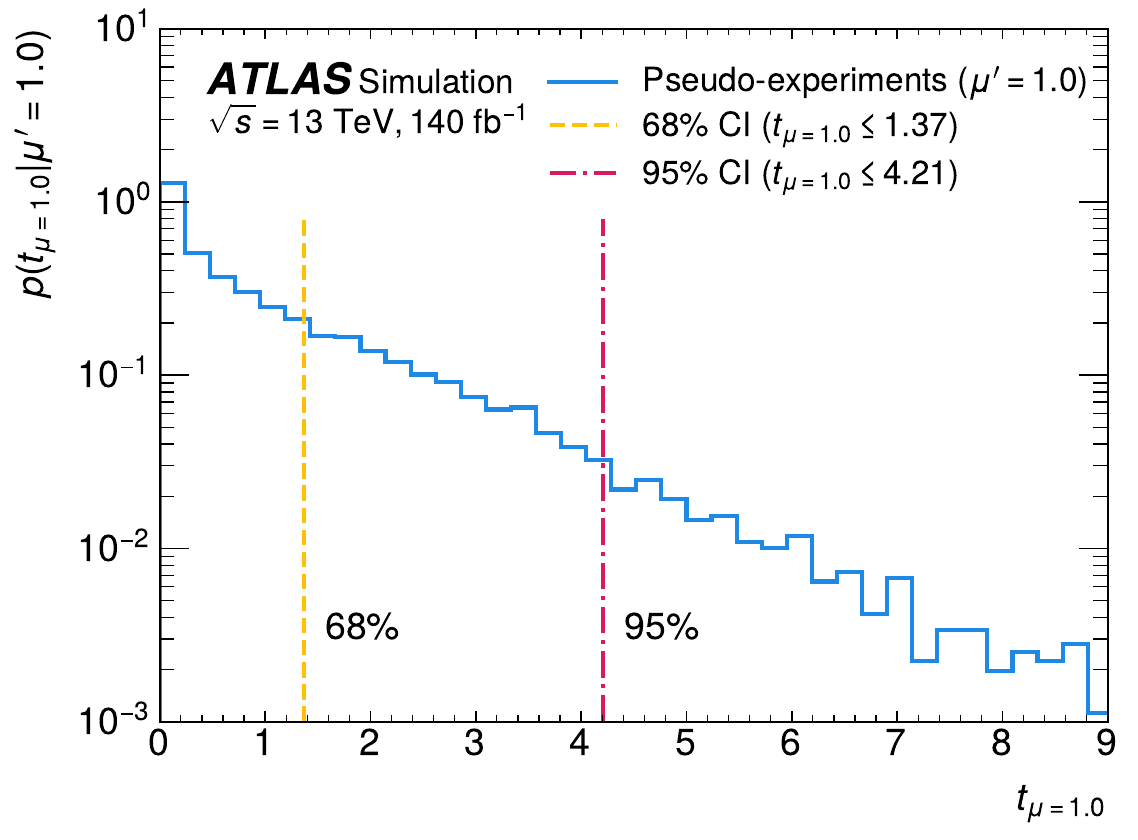}\label{fig:NeymanConstruction_t_mu_1p00_1}}
\caption{\protect\subref{fig:NeymanConstruction_t_mu_1p00_0} Distribution of the test statistic $t_{\mu=0.0}$ for the case $\mu^{\prime}=0.0$ and \protect\subref{fig:NeymanConstruction_t_mu_1p00_1} distribution of $t_{\mu=1.0}$ for the case with $\mu^{\prime}=1.0$. Each distribution is estimated with $15\,000$ pseudo-experiments. The confidence intervals (CI) are built using a Neyman construction by integrating up to 68.27\% (vertical dashed yellow line) and 95.45\% (vertical dash-dotted red line) of the distribution.}
\label{fig:NeymanConstruction_t_mu_1p00}
\end{figure}

\begin{figure}[tbp]
\vspace{0.1in}
\centering
\includegraphics[width=0.6\textwidth]{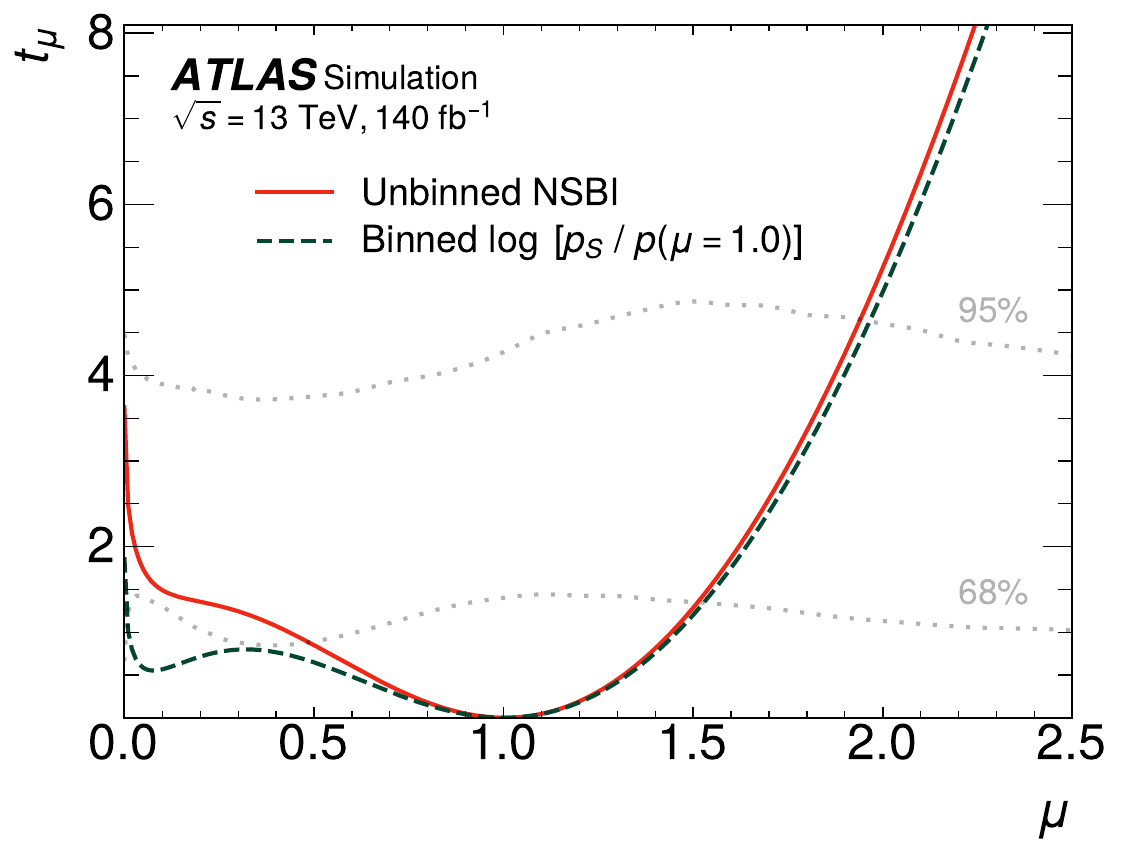}
\caption{A comparison of expected sensitivity of NSBI (solid red line) to a typical histogram-based (dashed green line) analysis, not including systematic uncertainties. The evaluation is performed on an Asimov dataset generated with $\mu=1$. The test statistic, the log-likelihood ratio $t_\mu$, is shown as a function of signal strength $\mu$. The 68\% and 95\% confidence intervals (CI) in dotted gray lines are determined using the Neyman construction.}
\label{fig:improvement_comparison}
\end{figure}

Once the pseudo-experiments are generated, the confidence intervals are built following the Neyman construction~\cite{1937RSPTA.236..333N}.
For the analysis described in \Sect{\ref{sec:ExampleUseCase}}, the distribution of $p(t_\mu | \mu)$, representing the test statistic $t_{\mu}$ for pseudo-experiments generated at a fixed value of $\mu$, is used to determine the one and two standard-deviation confidence intervals as functions of $\mu$. In each pseudo-experiment, the values of the AOs $a_k$ are sampled from the constraint density. The distribution of the test statistic values over many pseudo-experiments is shown in \Fig{\ref{fig:NeymanConstruction_t_mu_1p00}} with a $\mu$ of 1. This procedure is repeated over the range of $\mu$ to construct complete confidence bands as shown in \Fig{\ref{fig:improvement_comparison}}. The shapes of these bands deviate slightly from the asymptotic $\chi^2$ distribution in which the 68.27\% and 95.45\% confidence intervals would be defined exacly at $t_{\mu}=1$ and $t_{\mu}=4$. In the case of the off-shell Higgs production analysis, the deviation comes from the non-linear parametrization used in the off-shell Higgs boson production measurement~\cite{HIGG-2018-32},  and are not specifically a feature of NSBI\@.

The formalism discussed in this section lends itself to further tests for robustness on samples generated by shifting multiple NPs simultaneously and verifying that the confidence bands remain well-behaved in such scenarios. Such samples can be generated by a reweighting procedure similar to the one described in \Sect{\ref{subsec:NC_neg_weights}}, this time using the probability density ratio of \Eqn{\ref{eqn:factorisedGeneralForm}} that includes NPs,
\begin{equation}
w_i^{\text{rwt-ref}} \rightarrow w_i^{\text{Asimov}} (\mu, \alpha) = \frac{\nu(\mu, \alpha)}{\nu_\text{rwt-ref}} \; \frac{p(x_i|\mu, \alpha)}{p_{\text{rwt-ref}}(x_i)} \; w_i^{\text{rwt-ref}}.
\end{equation}

% End of text imported from the .//./sections/NeymanConstruction.tex input file

\section{Comparison of sensitivity}
\label{sec:resultComparison}

% The next lines are included from the .//./sections/ResultComparison.tex input file
This section shows the sensitivity of the NSBI method and the impact of systematic uncertainties in the result. The demonstration is performed for the simplified version of an off-shell Higgs boson signal strength measurement on simulated samples described in \Sect{\ref{sec:ExampleUseCase}} and considers a subset of the physics processes and systematic uncertainties described in \Sect{\ref{sec:example_sys_unc}}. The simplified analysis presented here contains only three processes and two sources of systematic uncertainties. A full analysis usually contains 10 -- 20 processes and 100 -- 200 independent systematic uncertainties. The simplified version allows the demonstration of the main features of the NSBI implementation presented in this paper without excessive computational costs.

\begin{figure}[tbp]
\vspace{0.1in}
\centering
\includegraphics[width=0.6\textwidth]{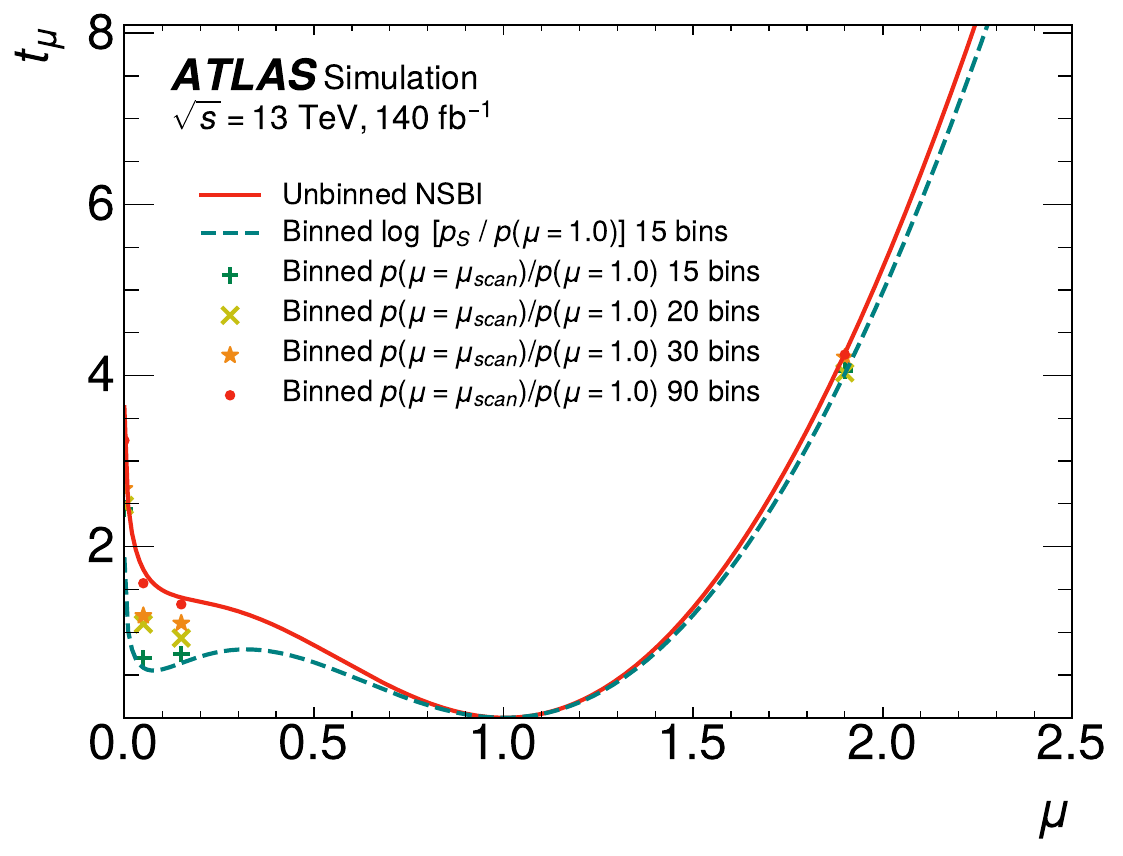}
\caption{A comparison of expected sensitivity from various analysis strategies using the log-likelihood ratio test statistic $t_\mu$, as a function of $\mu$. The evaluation is performed on an Asimov dataset generated with $\mu=1$. The solid red curve represents NSBI\@. The dashed green curve represents a typical histogram analysis that uses a fixed observable, $\log p_{\text{S}}/p(x|\mu=1)$, as a discriminant, with 15 bins. The markers show the sensitivity for various histogram analyses that use specific discriminants, $p(x_i|\mu)/p(x_i|\mu=1)$, for specific values of $\mu ( = 0.0 , 0.05, 0.15, 1.90)$, with 15 (green pluses), 20 (yellow crosses), 30 (orange stars) or 90 (red dots) bins. The improved sensitivity of the green plus markers over the green dashed curve, both using 15 bins, is due to the use of a parametrized observable.}
\label{fig:improvement_diagnostics}
\end{figure}
\subsection{Comparison to histogram-based methods}
The NSBI method is compared with two histogram-based analysis strategies on an Asimov simulated dataset, to show the gains due to the parametrized and unbinned nature of the method. The first histogram method employs a single observable, a discriminant between signal and full processes that is commonly used for LHC analyses,

\begin{equation}
\text{O}_\text{fixed} = \log\frac{p_\text{S}(x_i)}{p_\text{SBI}(x_i)}.
\end{equation}
Since this ratio is already estimated with ensembles for the NSBI method, no additional NNs need to be trained. This observable is subsequently used to construct a histogram (with 15 bins), and a Poisson likelihood fit is performed with it, analogous to what would be done in traditional analyses. The likelihood is used to construct a likelihood ratio test statistic, analogous to a traditional analysis using the same data. The improvement from NSBI relative to this approach can be seen in \Fig{\ref{fig:improvement_comparison}}.

To assess the power of the method exclusively due to the parametric approach of NSBI, its results are also compared to that of a parametrized but binned variant of NSBI. The second method uses an observable that is parametrized in $\mu$,
\begin{equation}
\text{O}_\mu = \frac{p(x_i|\mu)}{p(x_i|\mu=1)},
\end{equation}
that is subsequently binned and used to perform a Poisson likelihood fit. The log-likelihood ratio is computed for each value of $\mu$ using a histogram of the corresponding version of $\text{O}_\mu$, similar to the method described in Ref.~\cite{Baldi:2016fzo}. The improvement shown in \Fig{\ref{fig:improvement_diagnostics}} for $\text{O}_\mu$ over $\text{O}_\text{fixed}$ illustrates the power of a parametrized method. The traditional analysis (with the fixed observable) exhibits two prominent minima, which is typical in analyses with non-linear effects from, for example, quantum interference. However, the minimum at the incorrect value of $\mu$ is far less prominent for the analysis using a parametrized observable. Since the observable is optimized for each value of the parameter of interest, the method is able to more confidently reject the incorrect values of $\mu$. The further improvement of the full NSBI implementation is due to the unbinned nature of the method. As the number of bins increases, $\text{O}_\mu$ can approach the sensitivity of NSBI; however, this may introduce numerical instability, requiring careful bin width optimization, and make sufficiently fine binning untenable across the full range of $\mu$. If the number of bins in a histogram-based analysis is limited by the number of simulated events, then leveraging the power of unbinned fits may be desirable.

\begin{figure}[tbp]
\centering
\subfloat[]{\includegraphics[width=0.47\textwidth]{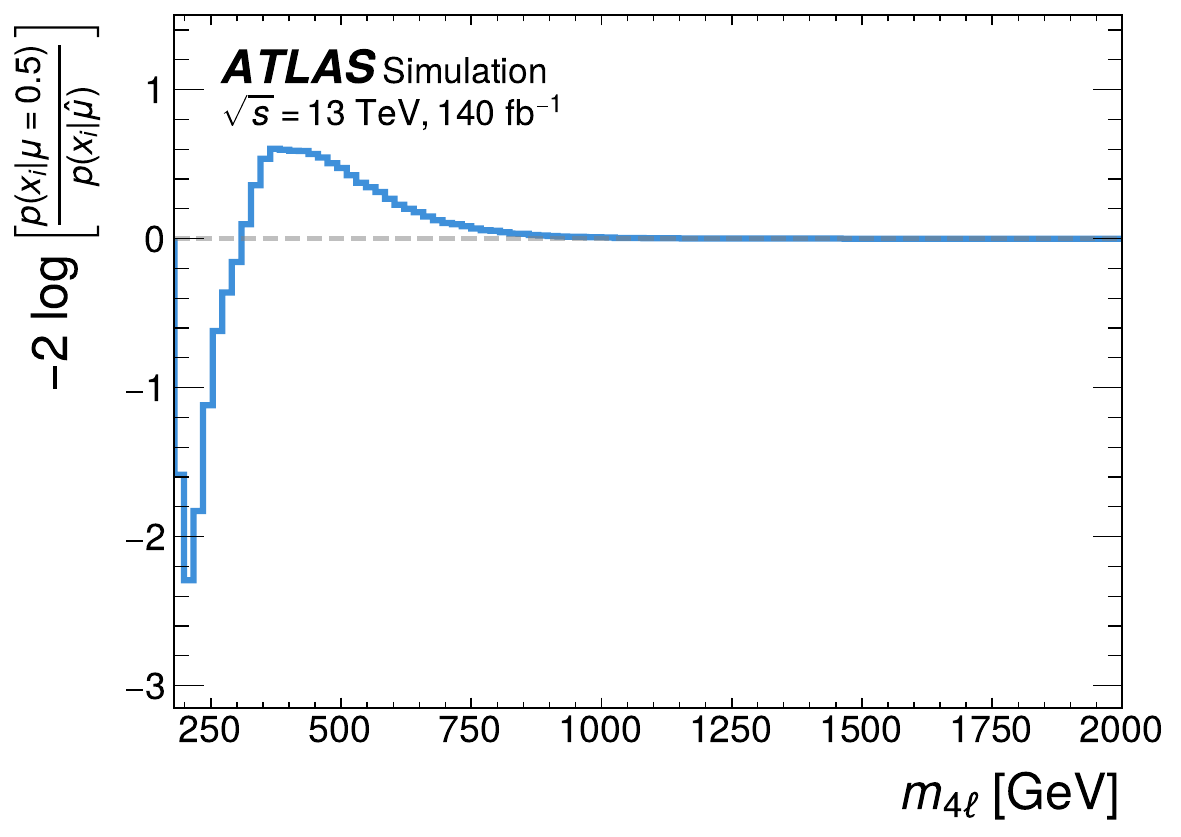}\label{fig:InterpretSensitivity_0p5}}\hfill
\subfloat[]{\includegraphics[width=0.47\textwidth]{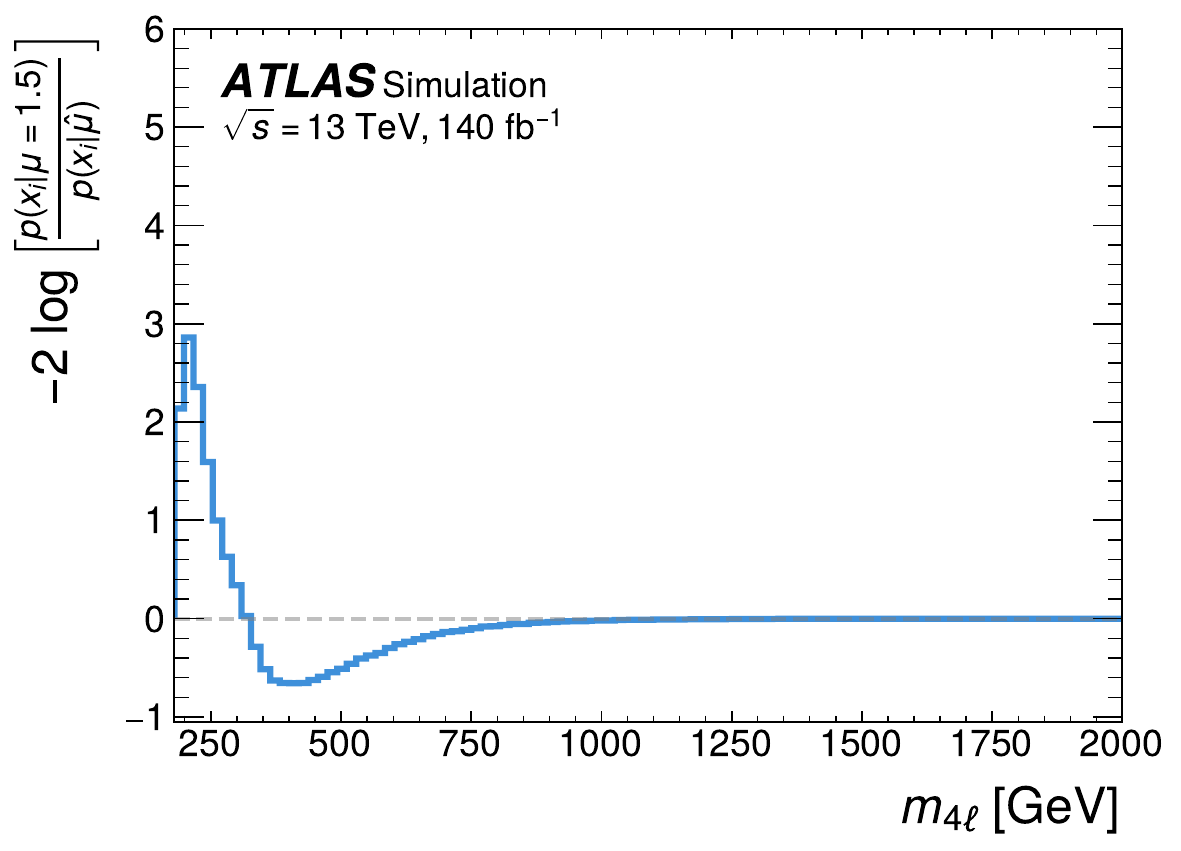}\label{fig:InterpretSensitivity_1p5}}
\caption{The sum of log density-ratios $-2\log(p(x_i|\mu^\prime)/p(x_i|\hat{\mu}))$ for events from an Asimov sample in bins of $m_{4\ell}$. The sum profile is shown for \protect\subref{fig:InterpretSensitivity_0p5} a hypothesis $\mu^\prime=0.5$ and \protect\subref{fig:InterpretSensitivity_1p5} a hypothesis $\mu^\prime=1.5$. This represents the per-event contribution to the test statistic for a given hypothesis, as a function of $m_{4\ell}$. Events in regions with the sum different than zero provide information to discriminate between the two hypotheses, $\mu^\prime$ and $\widehat{\mu}$, while regions with the sum close to zero do not impact the expected sensitivity of the result. The very high-mass region ($m_{4\ell}>750\,\GeV$) is equally consistent with both hypotheses and provides no additional sensitivity.}
\label{fig:InterpretSensitivity}
\end{figure}

An additional tool to interpret the results is shown in \Fig{\ref{fig:InterpretSensitivity}}, where the per-event contribution to the test statistic, $-2\log(p(x_i|\mu^\prime)/p(x_i|\widehat{\mu}))$, is summed over events from an Asimov sample in bins of $m_{4\ell}$. The profile is shown for two different hypotheses $\mu^\prime=0.5$ and $\mu^\prime=1.5$. Events in regions with the sum different than 0 provide information to discriminate between the two hypotheses, $\mu^\prime$ and $\widehat{\mu}$, while regions with the sum close to zero do not impact the expected sensitivity of the result. However, these one-dimensional distributions marginalize over the rest of the high-dimensional phase space. A region with sum close to zero may result from large cancellations in other dimensions, and a single distribution may not be enough to draw conclusions about a high-dimensional analysis.

\subsection{Impact of systematic uncertainties}

\begin{figure}[tbp]
\vspace{0.1in}
\centering
\includegraphics[width=0.66\textwidth]{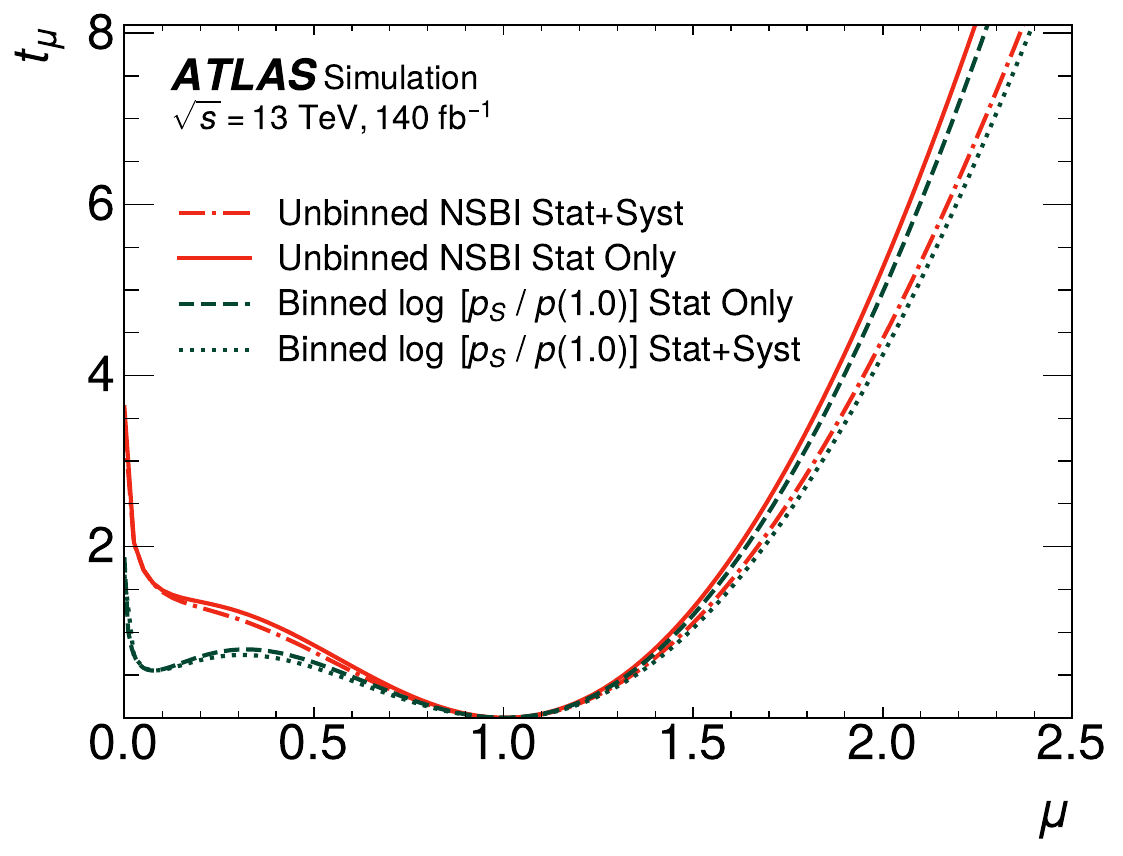}
\caption{Values of the log-likelihood test statistic $t_{\mu}$ as a function of signal strength $\mu$, representing only statistical uncertainties (solid red for NSBI, dashed green for histogram analysis), compared with the values of the profile log-likelihood ratio, representing both statistical and systematic uncertainties (dash-dotted red for NSBI, dotted green for histogram analysis), evaluated on Asimov data generated with $\mu=1$. The histogram analysis is performed with a fixed observable, $\log p_s/p(x|\mu=1)$. The two NPs in this study are described in \Sect{\ref{sec:example_sys_unc}}.}
\label{fig:profile_likelihood}
\end{figure}

The systematic uncertainties considered in this demonstration are described in \Sect{\ref{sec:example_sys_unc}}, and their impact is taken into account following the formalism developed in \Sect{\ref{sec:systematics}}. The $g_j(x_i,\alpha_k)$ term in \Eqn{\ref{eqn:factorisedGeneralForm}} accounts for the impact on the shape of the distributions and the $G_j(\alpha_k)$ term accounts for the impact on the inclusive rate. The interpolation functions used are described in \App{\ref{sec:ImplimentationAppendix}}. In the case of uncertainties that affect the inclusive rate, but not the shape of distributions, the term $g_j(x_i,\alpha_k)$ in \Eqn{\ref{eqn:factorisedGeneralForm}} is fixed to $1$ over the full range of $\alpha_k$. The profile (log-)likelihood is shown in \Fig{\ref{fig:profile_likelihood}} and compared with a histogram analysis using the $\text{O}_\text{fixed}$ observable. The systematic uncertainties reduce the sensitivity of the measurement, as is expected.

% End of text imported from the .//./sections/ResultComparison.tex input file

\FloatBarrier
\section{Conclusions and outlook}
\label{sec:conclusion}

% The next lines are included from the .//./sections/Conclusion.tex input file
While NSBI methods have drawn interest for their potential to dramatically improve the sensitivity of key analyses at the LHC, several open questions have remained regarding their application in a full-scale LHC analysis. This work develops the necessary tools and concepts required to have a complete statistical framework for NSBI at the LHC and addresses these open questions. The power and feasibility of the method are assessed through an example use case: a simplified measurement of the off-shell Higgs boson couplings in the four-lepton final states. This is an analysis with destructive quantum interference between the signal and background processes, which makes the likelihood model non-linear in the signal strength parameter and benefits from the power of NSBI methods. Comparisons with two histogram-based methods illustrate the gains from the unbinned and parametrized nature of the NSBI method. Since this demonstration was performed on a simplified version of the analysis that does not include all the relevant physics processes and systematic uncertainties, the expected sensitivity shown does not reflect the expected sensitivity of the full physics result.

The framework extends the standard statistical methodology employed at the LHC, transitioning to an unbinned, multi-dimensional setting, capable of accommodating a large number of systematic uncertainties. The paper also provides a list of diagnostics that can be used to understand and validate the performance of the neural network classifiers and describes a method to build a robust test statistic needed for hypothesis tests. It also describes the procedure to construct confidence intervals for unbinned analyses such as those using NSBI\@. Computational challenges in evaluating and analyzing the test statistic are overcome with the use of auto-differentiation techniques, which facilitate the profiling and the computation of pulls and impacts.

This method can be applied for parameter estimation in various frequentist statistical analyses, making optimal use of both the available data and simulated samples. It is particularly advantageous for analyses with non-linear likelihood models, large quantum interference, a small number of data events, or those requiring complex analysis observables. While ``optimal observables'' have previously been used to measure theory parameters in EFT analyses, these observables are close to optimal only for small regions of the parameter space, and often optimized for regions near the Standard Model. They typically do not account for detector effects. In contrast, NSBI is designed to achieve close to optimal sensitivity throughout the phase space under consideration, accounting for detector effects and systematic uncertainties.

Since this method inherits formalisms from the standard statistical methods used in LHC experiments, it also inherits the challenges. These include the challenge of model misspecification, for instance, if the simulation has systematic differences from data and the systematic uncertainties have not been well modeled. These issues can be diagnosed by inspecting pulls, impacts and other diagnostics in the same manner as in traditional histogram-based analyses. This method can also be used in conjunction with data-driven background estimates when the background simulations are not reliable and incorporate other techniques used in traditional analyses to mitigate systematic uncertainties. One potential technical challenge to an NSBI analysis over a histogram-based one is the need for enough training data to optimize precise probability density ratio estimators. These could be overcome by pre-training the networks first on larger datasets, such as fast simulated samples. Another technical challenge lies in the computational cost of training such a large number of neural networks. However, the increasing availability of large scientific computing facilities may mitigate this concern in the near future.

% End of text imported from the .//./sections/Conclusion.tex input file

%
\section*{Acknowledgements}

%

% The next lines are included from the .//acknowledgements/Acknowledgements.tex input file
%
%

%
%

We thank CERN for the very successful operation of the LHC and its injectors, as well as the support staff at
CERN and at our institutions worldwide without whom ATLAS could not be operated efficiently.

The crucial computing support from all WLCG partners is acknowledged gratefully, in particular from CERN, the ATLAS Tier-1 facilities at TRIUMF/SFU (Canada), NDGF (Denmark, Norway, Sweden), CC-IN2P3 (France), KIT/GridKA (Germany), INFN-CNAF (Italy), NL-T1 (Netherlands), PIC (Spain), RAL (UK) and BNL (USA), the Tier-2 facilities worldwide and large non-WLCG resource providers. Major contributors of computing resources are listed in Ref.~\cite{ATL-SOFT-PUB-2025-001}.

We gratefully acknowledge the support of ANPCyT, Argentina; YerPhI, Armenia; ARC, Australia; BMWFW and FWF, Austria; ANAS, Azerbaijan; CNPq and FAPESP, Brazil; NSERC, NRC and CFI, Canada; CERN; ANID, Chile; CAS, MOST and NSFC, China; Minciencias, Colombia; MEYS CR, Czech Republic; DNRF and DNSRC, Denmark; IN2P3-CNRS and CEA-DRF/IRFU, France; SRNSFG, Georgia; BMBF, HGF and MPG, Germany; GSRI, Greece; RGC and Hong Kong SAR, China; ICHEP and Academy of Sciences and Humanities, Israel; INFN, Italy; MEXT and JSPS, Japan; CNRST, Morocco; NWO, Netherlands; RCN, Norway; MNiSW, Poland; FCT, Portugal; MNE/IFA, Romania; MSTDI, Serbia; MSSR, Slovakia; ARIS and MVZI, Slovenia; DSI/NRF, South Africa; MICIU/AEI, Spain; SRC and Wallenberg Foundation, Sweden; SERI, SNSF and Cantons of Bern and Geneva, Switzerland; NSTC, Taipei; TENMAK, T\"urkiye; STFC/UKRI, United Kingdom; DOE and NSF, United States of America.

Individual groups and members have received support from BCKDF, CANARIE, CRC and DRAC, Canada; CERN-CZ, FORTE and PRIMUS, Czech Republic; COST, ERC, ERDF, Horizon 2020, ICSC-NextGenerationEU and Marie Sk{\l}odowska-Curie Actions, European Union; Investissements d'Avenir Labex, Investissements d'Avenir Idex and ANR, France; DFG and AvH Foundation, Germany; Herakleitos, Thales and Aristeia programmes co-financed by EU-ESF and the Greek NSRF, Greece; BSF-NSF and MINERVA, Israel; NCN and NAWA, Poland; La Caixa Banking Foundation, CERCA Programme Generalitat de Catalunya and PROMETEO and GenT Programmes Generalitat Valenciana, Spain; G\"{o}ran Gustafssons Stiftelse, Sweden; The Royal Society and Leverhulme Trust, United Kingdom.

In addition, individual members wish to acknowledge support from Armenia: Yerevan Physics Institute (FAPERJ); CERN: European Organization for Nuclear Research (CERN DOCT); Chile: Agencia Nacional de Investigaci\'on y Desarrollo (FONDECYT 1230812, FONDECYT 1230987, FONDECYT 1240864); China: Chinese Ministry of Science and Technology (MOST-2023YFA1605700, MOST-2023YFA1609300), National Natural Science Foundation of China (NSFC - 12175119, NSFC 12275265, NSFC-12075060); Czech Republic: Czech Science Foundation (GACR - 24-11373S), Ministry of Education Youth and Sports (ERC-CZ-LL2327, FORTE CZ.02.01.01/00/22\_008/0004632), PRIMUS Research Programme (PRIMUS/21/SCI/017); EU: H2020 European Research Council (ERC - 101002463); European Union: European Research Council (ERC - 948254, ERC 101089007, ERC, BARD, 101116429), Horizon 2020 Framework Programme (MUCCA - CHIST-ERA-19-XAI-00), European Union, Future Artificial Intelligence Research (FAIR-NextGenerationEU PE00000013), Italian Center for High Performance Computing, Big Data and Quantum Computing (ICSC, NextGenerationEU); France: Agence Nationale de la Recherche (ANR-20-CE31-0013, ANR-21-CE31-0013, ANR-21-CE31-0022, ANR-22-EDIR-0002); Germany: Baden-Württemberg Stiftung (BW Stiftung-Postdoc Eliteprogramme), Deutsche Forschungsgemeinschaft (DFG - 469666862, DFG - CR 312/5-2); Italy: Istituto Nazionale di Fisica Nucleare (ICSC, NextGenerationEU), Ministero dell'Università e della Ricerca (NextGenEU PRIN20223N7F8K M4C2.1.1); Japan: Japan Society for the Promotion of Science (JSPS KAKENHI JP22H01227, JSPS KAKENHI JP22H04944, JSPS KAKENHI JP22KK0227, JSPS KAKENHI JP23KK0245); Norway: Research Council of Norway (RCN-314472); Poland: Ministry of Science and Higher Education (IDUB AGH, POB8, D4 no 9722), Polish National Agency for Academic Exchange (PPN/PPO/2020/1/00002/U/00001), Polish National Science Centre (NCN 2021/42/E/ST2/00350, NCN OPUS 2023/51/B/ST2/02507, NCN OPUS nr 2022/47/B/ST2/03059, NCN UMO-2019/34/E/ST2/00393, NCN \& H2020 MSCA 945339, UMO-2020/37/B/ST2/01043, UMO-2021/40/C/ST2/00187, UMO-2022/47/O/ST2/00148, UMO-2023/49/B/ST2/04085, UMO-2023/51/B/ST2/00920); Portugal: Foundation for Science and Technology (FCT); Spain: Generalitat Valenciana (Artemisa, FEDER, IDIFEDER/2018/048), Ministry of Science and Innovation (MCIN \& NextGenEU PCI2022-135018-2, MICIN \& FEDER PID2021-125273NB, RYC2019-028510-I, RYC2020-030254-I, RYC2021-031273-I, RYC2022-038164-I); Sweden: Carl Trygger Foundation (Carl Trygger Foundation CTS 22:2312), Swedish Research Council (Swedish Research Council 2023-04654, VR 2018-00482, VR 2021-03651, VR 2022-03845, VR 2022-04683, VR 2023-03403), Knut and Alice Wallenberg Foundation (KAW 2018.0458, KAW 2019.0447, KAW 2022.0358); Switzerland: Swiss National Science Foundation (SNSF - PCEFP2\_194658); United Kingdom: Leverhulme Trust (Leverhulme Trust RPG-2020-004), Royal Society (NIF-R1-231091); United States of America: U.S. Department of Energy (ECA DE-AC02-76SF00515), Neubauer Family Foundation.

%
%

% End of text imported from the .//acknowledgements/Acknowledgements.tex input file

% The next lines are included from the .//sections/Acknowledgements_special.tex input file
This work started as part of the ATLAS Google Cloud Platform project~\cite{SOFT-2023-02}. Later stages of the work utilized computing resources at Southern Methodist University’s O’Donnell Data Science and Research Computing Institute and computing resources at the University of Massachusetts Amherst Research Computing.

% End of text imported from the .//sections/Acknowledgements_special.tex input file

%
%
%
%
%
%
%

%
\clearpage
\appendix
\part*{Appendix}
\addcontentsline{toc}{part}{Appendix}
\section{Interpolation function}
\label{sec:ImplimentationAppendix}

% The next lines are included from the .//./sections/Syst_appendix.tex input file
Section~\ref{sec:systematics} discusses the use of interpolation methods for systematic uncertainties. A common choice for the interpolation function to parametrize the impact of NPs at the LHC is~\cite{Cranmer:2014lly}
\begin{equation}
G_J(\va_k) = \begin{cases}
\left(\frac{\nu_J(\alpha_k^{(+)})}{\nu_J(\alpha_k^0)}\right)^{\alpha_k}& \alpha_k>1\\
1+\sum_{n=1}^6c_n\alpha_k^n& -1\leq\alpha_k\leq 1\\
\left(\frac{\nu_J(\alpha_k^{(-)})}{\nu_J(\alpha_k^0)}\right)^{-\alpha_k}& \alpha_k<-1\\
\end{cases},
\end{equation}
where the six coefficients $c_n$ of the polynomial in $\alpha_k$ are determined uniquely from the requirements that $G_J(\alpha_k)$ be continuous and its first and second derivatives be continuous at $\alpha_k=\pm 1$.
The same interpolation strategy and continuity requirements can be used to interpolate $g_J(x_i,\alpha_k)$,
\begin{equation}
g_J(x_i,\alpha_k) = \begin{cases}\displaystyle\left(g_J (x_i, \va_k^{(+)})\right)^{\alpha_k} & \alpha_k > 1\\\displaystyle
1+\sum_{n=1}^6 c_n\alpha^n_k & -1 \leq \alpha_k \leq 1\\\displaystyle
\left(g_J (x_i, \va_{k}^{(-)})\right)^{-\alpha_k} & \alpha_k < -1\end{cases}.
\end{equation}

% End of text imported from the .//./sections/Syst_appendix.tex input file

%
%
%
\printbibliography
\clearpage
% ATLAS Collaboration author list
% Reference date of SOFT-2023-01 is 2024-10-04
% Author list last updated on date 31-MAR-25
% Data extracted on 31-Mar-2025 for paper reference SOFT-2023-01
% at 11:48am
 
\begin{flushleft}
\hypersetup{urlcolor=black}
{\Large The ATLAS Collaboration}

\bigskip

\AtlasOrcid[0000-0002-6665-4934]{G.~Aad}$^\textrm{\scriptsize 104}$,
\AtlasOrcid[0000-0001-7616-1554]{E.~Aakvaag}$^\textrm{\scriptsize 17}$,
\AtlasOrcid[0000-0002-5888-2734]{B.~Abbott}$^\textrm{\scriptsize 123}$,
\AtlasOrcid[0000-0002-0287-5869]{S.~Abdelhameed}$^\textrm{\scriptsize 119a}$,
\AtlasOrcid[0000-0002-1002-1652]{K.~Abeling}$^\textrm{\scriptsize 56}$,
\AtlasOrcid[0000-0001-5763-2760]{N.J.~Abicht}$^\textrm{\scriptsize 50}$,
\AtlasOrcid[0000-0002-8496-9294]{S.H.~Abidi}$^\textrm{\scriptsize 30}$,
\AtlasOrcid[0009-0003-6578-220X]{M.~Aboelela}$^\textrm{\scriptsize 46}$,
\AtlasOrcid[0000-0002-9987-2292]{A.~Aboulhorma}$^\textrm{\scriptsize 36e}$,
\AtlasOrcid[0000-0001-5329-6640]{H.~Abramowicz}$^\textrm{\scriptsize 155}$
% ...
\bigskip
\\

% ...
$^{17}$Department for Physics and Technology, University of Bergen, Bergen; Norway.\\
% ...
$^{30}$Physics Department, Brookhaven National Laboratory, Upton NY; United States of America.\\
% ...
$^{36}$$^{(a)}$Facult\'e des Sciences Ain Chock, Universit\'e Hassan II de Casablanca;$^{(b)}$Facult\'{e} des Sciences, Universit\'{e} Ibn-Tofail, K\'{e}nitra;$^{(c)}$Facult\'e des Sciences Semlalia, Universit\'e Cadi Ayyad, LPHEA-Marrakech;$^{(d)}$LPMR, Facult\'e des Sciences, Universit\'e Mohamed Premier, Oujda;$^{(e)}$Facult\'e des sciences, Universit\'e Mohammed V, Rabat;$^{(f)}$Institute of Applied Physics, Mohammed VI Polytechnic University, Ben Guerir; Morocco.\\
% ...
$^{46}$Physics Department, Southern Methodist University, Dallas TX; United States of America.\\
% ...
$^{50}$Fakult\"{a}t Physik , Technische Universit{\"a}t Dortmund, Dortmund; Germany.\\
% ...
$^{56}$II. Physikalisches Institut, Georg-August-Universit\"{a}t G\"ottingen, G\"ottingen; Germany.\\
% ...
$^{104}$CPPM, Aix-Marseille Universit\'e, CNRS/IN2P3, Marseille; France.\\
% ...
$^{119}$$^{(a)}$New York University Abu Dhabi, Abu Dhabi;$^{(b)}$United Arab Emirates University, Al Ain; United Arab Emirates.\\
% ...
$^{123}$Homer L. Dodge Department of Physics and Astronomy, University of Oklahoma, Norman OK; United States of America.\\
% ...
$^{155}$Raymond and Beverly Sackler School of Physics and Astronomy, Tel Aviv University, Tel Aviv; Israel.\\
% ...
$^{a}$ Also Affiliated with an institute covered by a cooperation agreement with CERN.\\
$^{b}$ Also at An-Najah National University, Nablus; Palestine.\\
$^{c}$ Also at Borough of Manhattan Community College, City University of New York, New York NY; United States of America.\\
$^{d}$ Also at Center for High Energy Physics, Peking University; China.\\
$^{e}$ Also at Center for Interdisciplinary Research and Innovation (CIRI-AUTH), Thessaloniki; Greece.\\
$^{f}$ Also at CERN, Geneva; Switzerland.\\
% ...

\end{flushleft}

% Created with Glance <Atlas.Glance@cern.ch>

%

%
%
%
%

%
%
%
%

\end{document}